\documentclass{emulateapj}
\begin{document}

\newcommand{\chandra}{{\it Chandra}}
\newcommand{\swift}{{\it Swift}}
\newcommand{\xmm}{{\it XMM-Newton}}
\newcommand{\exosat}{{\it EXOSAT}}
\newcommand{\vaverageone}{$-1286\,\pm\,267$\,km\,s$^{-1}$}
\newcommand{\vaveragetwo}{$-771\,\pm\,65$\,km\,s$^{-1}$}

\title{The SSS phase of RS\,Ophiuchi observed with \chandra\ and \xmm\ I.\\
Data and preliminary Modeling}

\author{J.-U. Ness\altaffilmark{1}, S. Starrfield\altaffilmark{1}, A.P. Beardmore\altaffilmark{2}, M.F. Bode\altaffilmark{3}, J.J. Drake\altaffilmark{4}, A. Evans\altaffilmark{5}, R.D. Gehrz\altaffilmark{6}, M.R. Goad\altaffilmark{2}, R. Gonzalez-Riestra\altaffilmark{7}, P. Hauschildt\altaffilmark{8}, J. Krautter\altaffilmark{9}, T.J. O'Brien\altaffilmark{10}, J.P. Osborne\altaffilmark{2}, K.L. Page\altaffilmark{2}, R.A. Sch\"onrich\altaffilmark{11}, C.E. Woodward\altaffilmark{6}}

\altaffiltext{1}{School of Earth and Space Exploration, Arizona
State University, Tempe, AZ 85287-1404, USA: [Jan-Uwe.Ness, sumner.starrfield]@asu.edu}
\altaffiltext{2}{Department of Physics \& Astronomy, University of Leicester, Leicester, LE1 7RH, UK}
\altaffiltext{3}{Astrophysics Research Institute, Liverpool John Moores University, Birkenhead, CH41 1LD, UK}
\altaffiltext{4}{Harvard-Smithsonian Center for Astrophysics, 60
Garden Street, Cambridge, MA 02138, USA}
\altaffiltext{5}{Astrophysics Group, Keele University, Keele, Staffordshire, ST5 5BG, UK}
\altaffiltext{6}{Department of Astronomy, School of Physics \& Astronomy, 116 Church Street S.E., University of Minnesota, Minneapolis, MN 55455, USA}
\altaffiltext{7}{XMM Science Operations Centre, ESAC, P.O. Box 50727,
28080 Madrid, Spain}
\altaffiltext{8}{Hamburger Sternwarte, Gojenbergsweg 112, 21029
Hamburg, Germany}
\altaffiltext{9}{Landessternwarte K\"onigstuhl, 69117
Heidelberg, Germany}
\altaffiltext{10}{Jodrell Bank Observatory, School of Physics \& Astronomy, University of Manchester, Macclesfield, SK11 9DL, UK}
\altaffiltext{11}{Universit\"ats-Sternwarte der
Ludwig-Maximilians-Universit\"at,
Scheinerstr. 1, 81679 M\"unchen,
Germany}

\journalinfo{accepted 2007, May 8}

\begin{abstract}
The phase of Super-Soft-Source (SSS) emission of the sixth recorded
outburst of the recurrent nova RS\,Oph was observed twice with
\chandra\ and once with \xmm. The observations were taken on days 39.7
(\chandra), 54.0 (\xmm), and 66.9 (\chandra) after outburst. We confirm
a $\sim 35$-sec period on day 54.0 and found that it originates from
the SSS emission and not from the shock. We discus the bound-free
absorption by neutral elements in the line of sight, resonance absorption
lines plus self-absorbed emission line components, collisionally
excited emission lines from the shock, He-like intersystem lines, and
spectral changes during an episode of high-amplitude variability.
We find a decrease of the oxygen K-shell absorption edge that can
be explained by photoionization of oxygen. The absorption component
has average velocities of \vaverageone\ on day 39.7 and of
\vaveragetwo\ on day 66.9. The wavelengths of the emission line
components are consistent with their rest wavelengths as confirmed
by measurements of non-self absorbed He-like intersystem lines.
We have evidence that these lines originate from the shock rather
than the outer layers of the outflow and may be photoexcited in
addition to collisional excitations. We found collisionally excited
emission lines that are fading at wavelengths shorter than
15\,\AA\ that originate from the radiatively cooling shock.
 On day 39.5 we find a systematic blue shift of
$-526\,\pm\,114$\,km\,s$^{-1}$ from these lines.
We found anomalous He-like f/i ratios which indicates either high
densities or significant UV radiation near the plasma where the
emission lines are formed.
 During the phase of strong variability the spectral
hardness light curve overlies the total light curve when
shifted by 1000\,sec. This can be explained by photoionization of
neutral oxygen in the line of sight if the densities of order
$10^{10}-10^{11}$\,cm$^{-3}$.
\end{abstract}

\keywords{stars: novae, stars: individual (RS\,Oph) ---  X-rays: stars}

\section{Introduction}

RS\,Oph is a Symbiotic Recurrent Nova (RN) that had recorded
outbursts in 1898, 1933, 1958, 1967, 1985, and 2006 \citep[February
$12.83=$\,day\,0;][]{rsophdiscovery}. In previous outbursts RS\,Oph was
well-studied in the optical, and in 1985 it was followed extensively in
ultraviolet \citep[][and references therein]{shore96}, radio
\citep{padin85,hje86}, and X-rays \citep[six pointings by \exosat\
from day 55 to day 251 of the outburst,][]{mason87,obrien92}.

RS\,Oph is a member of a small class of Cataclysmic Variables (CVs) in
which a white dwarf (WD) orbits a red giant secondary (M2III).
The orbital period is $455.72\,\pm\,0.83$\,days and, assuming a WD near
the Chandrasekhar
Limit and a low inclination ($i<40^{o}$), the red giant mass is of order
0.5\,M$_\odot$ \citep{dobrKen94}. The distance to RS\,Oph has been well
determined by a variety of methods as $1.6\,\pm\,0.3$\,kpc \citep{bods87}.
The interstellar absorbing column,
N$_{\rm H} = (2.4\,\pm\,0.6)\times 10^{21}$\,cm$^{-2}$, was determined from
H\,{\sc i} 21\,cm measurements \citep{hje86} and is consistent with the
visual extinction ($E(B-V) = 0.73\pm0.1$) determined from IUE observations
in 1985 \citep{sni87}.

\setcounter{footnote}{1}
\renewcommand{\thefootnote}{\fnsymbol{footnote}}
Just as in Classical Novae (CNe\footnote{The companion of the WD
is generally a main-sequence star instead of a giant.}), accretion of
material from the secondary causes an explosion on the WD surface.
However, in RNe the rate of mass accretion and the WD mass
are sufficiently large for outbursts to repeat on human timescales.
(The other members of this class are T\,CrB, V745\,Sco, and V3890\,Sgr).

A RN evolves analogously to a CN, the major difference between the two
types of outburst arises from the presence of the red giant in the
binary system which completely changes the environment around the WD.
The evolution of the X-rays in RS\,Oph can be characterized by four phases:
\begin{itemize}
\item The explosion ejects material into its surroundings and produces
a strong shock moving into the wind and outer atmosphere of the red
giant and backwards into the ejecta
\citep[e.g.,][]{bode06,obrien06,sokoloski06}. The strength of the
shock depends on the kinetic energy of the
ejecta and the density of the medium into which the ejecta run.
\item While nuclear burning continues on the WD, the bolometric
luminosity is approximately constant \citep{gallcode74}.
Depending on the opacity due to electron scattering within the
expanding shell, the peak of the observed spectrum gradually shifts
from low energies (optical and UV) to soft X-rays \citep{gallcode74}.
In CNe this happens after a few weeks to months
\citep[e.g., V4743\,Sgr;][]{v4743}, and the spectrum resembles that of
the class of Super Soft X-ray Binary Sources \citep[SSS][]{kahab}. This
phase is therefore called the SSS phase.
\item After the end of nuclear burning on the WD, the SSS emission
decreases and RS\,Oph enters a phase in which a recombining plasma,
exhibiting emission lines from radiatively excited states, appears.
\item The final stage is that of the ejected material radiatively
cooling. Collisional bound-bound excitations are balanced by
radiative de-excitations, giving rise to emission lines (also called
the coronal approximation). In this way kinetic energy is effectively
converted into radiation.
\end{itemize}

The first few weeks of the evolution of the X-ray emitting blast wave
in RS\,Oph were studied by \cite{sokoloski06} using the Rossi X-ray
Timing Explorer (RXTE) and \cite{bode06} using \swift. The shock wave was
also extensively studied at radio wavelengths by Very Long Baseline
Interferometry (VLBI) and MERLIN imaging observations
\citep{obrien06}. From these observations, those obtained with the
Hubble Space Telescope \citep{bode07} and with ground-based infrared
interferometers \citep[VLTI+AMBER][]{Chesneau07}, asymmetries and multiple
emission components have clearly been established that show that there is
jet-like ejection in addition to shell-like ejection of material.
In the 2006 outburst \chandra\ obtained one, and
\xmm\ obtained two observations of this phase
\citep[][Ness in prep.]{rsoph_iau1,drake06,2006IAUC.8695....2G}.

 The SSS phase commenced about 30 days into the outburst \citep{atel770}.
Early in the evolution, the energy output of a nova explosion
happens primarily at optical and UV wavelengths, because the high-energy
radiation produced by nuclear burning on the surface of the WD
will be scattered within the surrounding shell, leaving the atmosphere
as lower-energy radiation. When the surrounding shell becomes thinner
as a consequence of the expansion, the optical brightness decreases
while the X-ray brightness increases and originates from deeper within
the outflow. X-ray observations during this phase allow studies of the
plasma regions at the radial distance from the WD within the outflow
where the optical depth $\tau\approx 1$. The spectrum is a hot stellar
atmosphere whose peak energy and shape are dominated by the temperature
and by
interstellar and circumstellar absorption. In CNe, additional absorption
lines have been observed \citep[e.g., V4743\,Sgr,][]{v4743} which
originate from highly ionized species and are shifted and broadened
according to the dynamics of the observed plasma.

X-ray observations of the 1985 outburst taken with \exosat\ after day 55
\citep{mason87} were modeled by \cite{obrien92} assuming the emission
to originate from the shocked wind. A direct comparison of the
shapes of the light curves taken with \exosat\ and with \swift\
in 2006 \citep{osborne06} show a remarkable resemblance, implying that
the spectra obtained in 1985 (at least in the case of the LE of
\exosat) were actually dominated by the SSS emission \citep{ness_suzaku}.

We present three grating observations taken during the SSS phase.
Two observations were taken with \chandra\ \citep{rsoph_iau2} and one with
\xmm. We describe our method of extraction of the X-ray light curves and
X-ray spectra in Sects.~\ref{lcsect} and \ref{spec}, respectively. We
present our spectral analysis in Sect.~\ref{anal}, focusing on the
broad-band continuum spectrum in Sect.~\ref{continuum}, the resonance
absorption- and emission lines in Sect.~\ref{pcygsect}, emission lines
originating from the shock in Sect.~\ref{elinesect}, He-like
intersystem lines in Sect.~\ref{hesect}, and the spectral changes
during brightness variations detected during the first observation
in Sect.~\ref{variability}.
We discuss various possible interpretations in Sect.~\ref{disc}
and summarize our conclusions in Sect.~\ref{concl}.

\section{Observations}

\begin{table}[!ht]
\begin{flushleft}
\renewcommand{\arraystretch}{1.1}
\caption{\label{tab1}Grating observations during SSS phase}
\begin{tabular}{lrllrl}
\multicolumn{2}{l}{2006 Date \hfill day$^{[a]}$}& Mission & Grating & ObsID & exp. time\\
start--stop & & & /detector & & (net)\\
\hline
March 24, 12:25:22 & 39.69 & \chandra & LETGS & 7296 & 10.0\,ksec\\
\multicolumn{2}{l}{--March 24, 15:38:20 \hfill 39.82} & & /HRC & &\\
April 07, 21:04:52 & 54.05 & {\it XMM} & \multicolumn{2}{l}{RGS1\hfill 0410180301} & 9.8\,ksec\\
\multicolumn{2}{l}{--April 08, 02:20:04 \hfill 54.27} & & RGS2& & 18.6\,ksec\\
April 20, 17:23:48 & 66.89 & \chandra & LETGS & 7297 & 6.5\,ksec\\
\multicolumn{2}{l}{--April 20, 20:27:57 \hfill 67.02} & & /HRC & &\\
\hline
\end{tabular}

$^{[a]}$after outburst (2006 Feb 12.83UT)
\renewcommand{\arraystretch}{1}
\end{flushleft}
\end{table}

During the SSS phase that started after day 26 (2006 March 10), three
grating observations were carried out by \chandra\ \citep{chandra} and \xmm\
\citep{xmm}. We summarize the dates of each observation, the respective days
after the outburst, instrumental setup, observation identification numbers
(ObsID), and the net exposure times in Table~\ref{tab1}. The first and the
last observations were
taken by \chandra\ with net exposure times of 10.0\,ksec and 6.5\,ksec,
respectively. As a consequence of the brightness of the source, telemetry
saturation (full buffer all the time) occurred, which reduced the effective
exposure times compared to the on-source times. Provisions were taken to
ensure that these
reduced spectra were correctly calibrated. We used the LETGS/HRC combination
\citep[Low Energy Transmission Grating/High Resolution Camera:][]{letg,hrc},
which is an imaging dispersion spectrometer. Higher dispersion orders are
not filtered out, but in our case contamination of the first-order spectrum
by higher-order photons is negligible because the SSS spectrum is
sufficiently narrow as to prevent an overlap of the dispersion orders.

The second grating observation during the SSS phase was taken with \xmm\
(XMM) on day 54 after outburst (2006 April 7). XMM carries five X-ray
instruments, three low-resolution CCD detectors and two gratings
\citep[Reflection Grating Spectrometers RGS1 and RGS2:][]{rgs}. Due to
the unexpected brightness of the target, the allocated telemetry
rates were not sufficient to handle the large event rate. All instruments
were affected by telemetry saturation. This implies that the effective
exposure times were shorter than the on-target times. For the RGS2, the
net exposure time was 18.6\,ksec but only 9.8\,ksec in the RGS1 (total
on-target time was 18.9\,ksec). Since the telemetry losses do not depend
on the energy of the photons, they have no effect on the extracted
spectrum, except for the shorter net exposure time. However, the light
curves were strongly affected, since telemetry losses cannot be recovered.
The X-rays dispersed into the spectrum by reflecting off the RGS are recorded
with a strip of CCD detectors, and the wavelength range is 5--38\,\AA.
Unfortunately, two chips within these CCD arrays failed early in the mission,
and those portions of the dispersed spectrum that should be recorded by
these chips are lost. The wavelengths affected by the chip failure range
from 10.5--13.8\,\AA\ in the RGS1 and 20--24\,\AA\ in the RGS2. Since the
peak emission of our SSS spectrum ranges between 15\,\AA\ and $\sim
30$\,\AA\ (see Sect.~\ref{spec} and Fig.~\ref{cmp_sss_rgs}), only the RGS1
gives us sufficient spectral information, and we concentrate primarily on
the RGS1 for spectroscopy. We use the RGS2 for light curve analyses.
Since the source spectrum has its peak in the middle of the RGS2 chip gap
(Fig.~\ref{cmp_sss_rgs}), the RGS2 collects fewer photons and is less
affected by telemetry losses. This may be the only example where the chip
gap is actually of use, providing us with the only useful light curve of
this observation.

We present our extraction of the light curves in Sect.~\ref{lcsect} and that
of the spectra in Sect.~\ref{spec}. We used standard tools provided by the
mission-specific software packages SAS (Science Analsis Software, version
7.0) and CIAO (Chandra Interactive Analysis of Observations, version 3.3.0.1).

\subsection{Extraction and analysis of light curves}
\label{lcsect}

\begin{figure}[!ht]
\resizebox{\hsize}{!}{\includegraphics{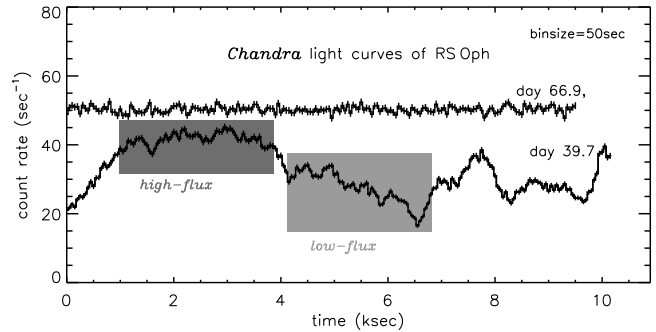}}
\caption{\label{lc}Light curves of the \chandra\ LETG observations, days 39.6 and
66.9, in time bins of 50\,sec. The light curves have been extracted from the
dispersed photons covering a wavelength range $\sim 6-50$\,\AA. For the first
observation we extracted separate spectra from the time intervals marked with
different shadings (see Fig.~\ref{tres}).}
\end{figure}
\chandra:
For the extraction of \chandra\ light curves we used the CIAO tool
{\tt lightcurve}, which extracts all photons within previously defined
source and background extraction regions on the detector. As source
extraction regions we chose two
polygons around the streaks of dispersed photons including both dispersion
directions and we used only the middle chip (i.e., $\lambda <50$\,\AA). The
background was extracted from adjacent regions and subtracted.

\begin{figure}[!ht]
\resizebox{\hsize}{!}{\includegraphics{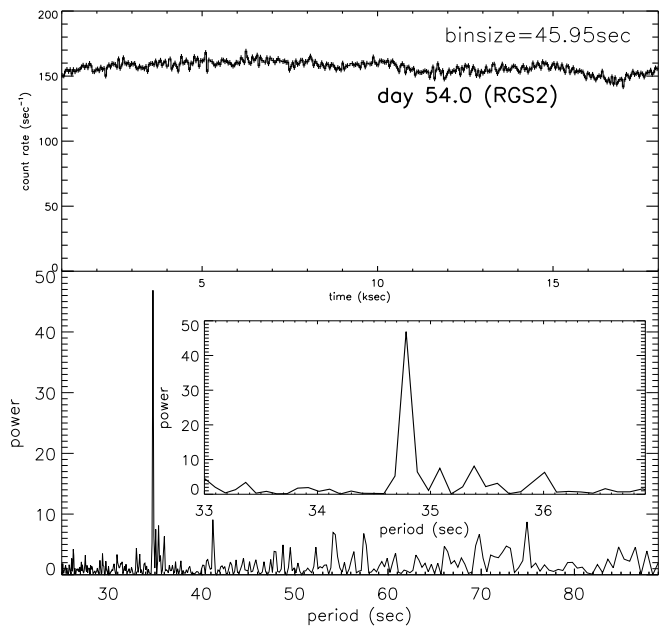}}
\caption{\label{lc_rgs}RGS2 light curve (day 54.0) in 45.947-sec time bins
(10 times readout time, top) and periodogram (bottom).
A $\sim 35$-sec periodicity is found in
the RGS2 light curve that confirms observations of a similar period with
\swift\ reported by \cite{atel770}.}
\end{figure}
\xmm:
For the extraction of \xmm\ RGS2 light curve we used the SAS tool {\tt evselect},
which extracts all dispersed photons within standard source- and background 
extraction regions. We extracted the RGS2 light
curve in four different bin sizes in time, all chosen to be multiples of
the readout time (4.5947\,sec) in order to search for periodicity.

In Fig.~\ref{lc} we show the \chandra\ light curves in 50-sec time bins in units
of ksec after the start of each observation. In the observation of day 39.7,
the count rate increases rapidly by a factor of two within only
1000\,seconds, remaining at a high count rate for about 3000\,seconds, then dips
down (a little bit lower than the start count rate), only to rise again with
an even steeper slope towards the end of the observation. In Fig.~\ref{lc_rgs}
we show the RGS2 light curve extract for day 54.0, which is much flatter than
that on day 39.7, but is not as flat as on day 66.9. For the latter light
curve we investigated the possibility of rapid variability, however, a
simple model assuming a constant emission level reproduces the measured
light curve with a reduced $\chi^2=0.94$, thus models including any
kind of variability cannot improve the fit by more than 68.3\%.

We computed a power spectrum for all light curves, and for days 39.7 and
66.9 we find no significant periodic variations. In order to search for
periods shorter than 50\,sec, we also extracted the \chandra\ light curves
in smaller time bins, but did not detect any periods longer than 2\,sec.
In the bottom panel of Fig.~\ref{lc_rgs} we show the result from the
period search of the RGS2 light curve. In this observation we find a
34.8-sec period that is consistent with the $\sim 35$-sec period reported
by \cite{atel770}. We also found a period of $\sim 5.3$\,sec, which is
the beat period of the instrument readout time (4.5947\,sec) and the
34.8-sec period. We extracted the light curves from each chip and found
this period only on those chips that recorded the SSS spectrum.

Our re-analysis of the \swift\ light curves obtained on days $\approx 39.7$,
54.0, and 66.9 indicated that the $\sim 35$\,sec periodicity is only
present on day 54.0. \cite{atel801} reported that this periodic
variability was not always present, disappearing entirely after day 63 … 
(2006, April 17).


We compared the emission levels in the grating spectra with those
of \swift\ at the same times. The first observation was taken during
the phase of highly variable soft X-ray flux reported in
\cite{atel764,atel770}. This phase lasted from days 29 to 46
\citep{atel801} and is thus coincident with the highly variable light
curve taken with \chandra\ on day 39.7. Since \swift\ light curves
are never continuous over more than at most 2\,ksec due to its orbit,
it is the \chandra\ light curve that demonstrates that the source was
actually variable on time scales shorter than 0.1 days.

Close inspection of the \swift\ monitoring light curve
\citep{osborne06} revealed that the
first \chandra\ observation (day 39.7) was taken at a time when
the \swift\ count rate was close to one of the minima during the
variability phase, while the second \chandra\ observation
(day 66.9) was taken during the decline shortly after the peak
emission level had been reached. This explains the higher
\chandra\ count rate on day 66.9 compared to day 39.7.
The \xmm\ observation was carried out during a phase of higher emission.
We compared the relative emission levels between days 39.7,
54.0 and 66.7 by computing the photon fluxes by integration of the
grating spectra (see Sect.~\ref{spec}). Within the uncertainties
of the flux variations on day 39.7, the changes are consistent
with the changes of the corresponding \swift\ count rates.

\subsection{Extraction of spectra}
\label{spec}

\begin{figure}[!ht]
\resizebox{\hsize}{!}{\includegraphics{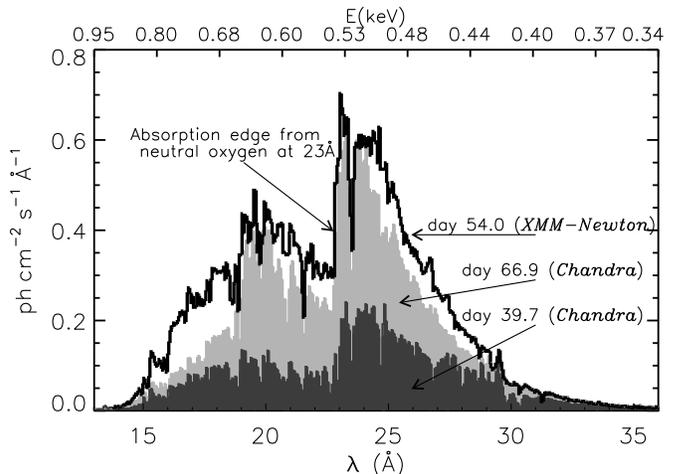}}

\caption{\label{cmp_sss_rgs}
Comparison of all three grating spectra in units of photon fluxes obtained by
dividing count rates by effective areas. The \chandra\ spectra are shown with
shading and the RGS1 spectrum is a thick solid histogram line.}
\end{figure}

Grating spectra are extracted on an equidistant wavelength grid (in units
of \AA), and we use wavelength units throughout this paper. We extracted
the spectra with the \chandra\ CIAO tool {\tt tgextract} for the LETGS
observations and the {\it XMM} SAS tool {\tt rgsproc} for the RGS
observation. These routines place standardized extraction regions over the
dispersed photons on the detector and are optimized to maximize the ratio
of collected source counts to the included background. The mirror point
spread function is translated into an instrumental line profile that is
approximately Lorentzian for the RGS and more Gaussian-like for \chandra\
LETG (both FWHM$\sim 0.055$\,\AA\ at all wavelengths, corresponding to
500--1100\,km\,s$^{-1}$ from 30\,\AA\ to 10\,\AA, respectively).
Velocities below these values that contribute to line broadening are
difficult to determine. However, velocities from line shifts can be
determined accurately (see Table~\ref{tab_profiles} in
Sect.~\ref{pcygsect}). Using standard tools to calculate the effective
areas (in cm$^2$) and then dividing a count rate by the effective area
at the corresponding wavelength results in a photon flux that is
sufficiently independent of the instrument (RGS vs. LETGS). Thus, the
photon flux spectra from the RGS1 and the LETGS can be compared
(see Fig.~\ref{cmp_sss_rgs}).

\chandra:
The dispersed photons are recorded in two streaks in opposite directions
from the zero-th order, delivering two independent spectra. We co-added
these two spectra for best signal to noise. In addition, independent
analyses of the spectra can be carried out for consistency checks. In
Fig.~\ref{cmp_sss_rgs} we present the two \chandra\ observations (light
and dark shadings) along with the {\it XMM} observation (solid line).

\xmm:
 RGS data were processed with the SAS task {\tt rgsproc} in SASv7.0
up to the creation of the merged, filtered event file. This file was 
manipulated to correct for pileup. Given the high flux of the source,
spectra were extracted 
from the full field of view instead of using standard extraction regions,
and no background subtraction was applied. The separation
of spectral orders was accomplished by using the energy resolution of the CCDs.
Pileup occurs when two or more events arrive at the same 
(or neighboring) pixel during the same readout frame. These photons are 
registered as a single event with an energy that is the sum of the energies of the
individual events. In our case, pileup results in events with CCD-measured
energies $n\times E$ (where $n=1,2,3,4$) while the photons are dispersed
according to their individual wavelengths. Due to this pile-up, counts occur in
the energy-dispersion plane in regions normally associated with higher orders.
There is no ambiguity between pile-up and higher order dispersion
due to the narrow spectral range of the super-soft continuum spectrum.
We were thus able to reconstruct the true spectrum before pile-up by adding those
events in the energy-dispersion plane back into the first order spectrum that
resulted from pile-up in the first-order spectrum. The rates of the new
first-order spectra are about 30\,percent higher than before this correction.

For a qualitative comparison of the spectra from the different missions we
converted all three count rate spectra into photon flux spectra. We do not
correct for the redistribution matrix, so the line profiles will still
depend on the individual instrumental point spread function (PSF).
However, the shape of the continuum is not affected by the instrumental
PSF, as it is very small compared to the width of the continuum.

In Fig.~\ref{cmp_sss_rgs} we compare the photon flux spectra of the three
grating observations. The integrated fluxes relative to each other are
consistent with the relative count rates \swift\ measured at the respective
times. All spectra show continuum emission over the same wavelength range
with similar shapes. At 22.83\,\AA\ there is a strong absorption edge in all
spectra that is clearly non-instrumental and originates from O\,{\sc i}
(K-shell ionization). Also, an expected narrow $1s-2p$ absorption line
at 23.5\,\AA\ \citep{paerels01_abs} from atomic oxygen can be identified
in all three spectra. At $\sim 29$\,\AA\, there are emission line features
in all spectra that we attribute to N\,{\sc vi} (see Sect.~\ref{hesect}).
Strong absorption
lines can be identified in all three spectra (see Sect.~\ref{pcygsect}).

\section{Spectral Analysis}
\label{anal}

\subsection{Continuum emission and broad-band absorption}
\label{continuum}

 In order to understand the cause of the brightness changes we
computed a series of blackbody models and found reasonable
agreement with the measured spectra for temperature ranges
$(630-830)\times10^3$\,K on day 39.7,
$(650-710)\times10^3$\,K on day 54.0, and $(590-720)\times10^3$\,K
on day 66.9. The bolometric luminosities, $\log(L_{\rm bol})$, of
these models are between 37.3 and 38.5 for day 39.7, 38.5 and 38.9
for day 54.0, and 38.2 and 39.3 for day 66.9. We used a standard
model for interstellar plus circumstellar absorption. The value
of hydrogen column density N$_{\rm H}$ was greater than the
interstellar value (N$_{\rm H}=2.4\times10^{21}$\,cm$^{-1}$)
for all models (more below).
However, the values of reduced $\chi^2$ ($\chi^2_{\rm red}$) are
greater than 20 for all models, and no secure conclusions can be
drawn from these numbers. In particular, we don't claim from these
models that Super-Eddington luminosities occurred. With these
limitations we are not able to explain the
brightness changes in terms of temperature or luminosity.

 As a first approach to characterize the shape of the continuum, we
computed hardness ratios HR=(H-S)/(H+S) from the three spectra with
H and S denoting the fluxes extracted from within the wavelength ranges
15--23\,\AA\ and 23--30\,\AA, respectively. We found values of
-0.06, +0.12, and +0.03 with 4\%, 1\%, and 8\% uncertainties for
days 39.7, 54.04, and 66.9, respectively. Changes in hardness generally
imply a change in temperature, but changes in broad-band absorption
(bound-free transitions) by elements in the line of sight can also lead
to changes in spectral hardness. We estimated the optical depth of the
O\,{\sc i} absorption edge at 22.83\,\AA\ from the intensities of
the continuum,
$\tau=\ln({\rm cont}_{\lambda<22.8}/{\rm cont}_{\lambda>22.8})$,
 and found $\tau=0.9,\ 0.5,$ and $0.7$ for days
39.7, 54.0, and 66.9, respectively. This implies that the larger
hardness ratio on day 54 was caused by reduced O\,{\sc i} within
the material in the line of sight.

In order to investigate this effect closer, we use the blackbody
models from above as continua and apply an absorption model that
accounts for variations of the neutral oxygen abundance. We optimized
the blackbody parameters temperature and bolometric luminosity
simultaneously with the parameters of the absorption model.

The continuum spectrum produced by the WD atmosphere has to pass
through the circumstellar and the interstellar material in the line
of sight. The former may be partially ionized while the latter consists of
neutral elements of solar composition. The bound-free absorption imposed
by this material leads to element-specific absorption edges that are
detectable with the LETGS \citep[e.g.,][]{paerels01_abs}.
The depth of these edges depends on the
elemental composition and absorption cross sections, e.g. for K-shell
absorption by neutral oxygen. Absorption edges from higher ionization stages
also depend on the fractional number density of the ion. We parameterize
bound-free absorption in the line of sight by the column density of neutral
hydrogen, N$_{\rm H}$. While the column density of the interstellar material
is assumed constant in all observations, it may vary with time in the
circumstellar material. We, therefore, split the absorption into two
terms, one term with fixed N$_{\rm H}$ and one with variable N$_{\rm H}$,
both with solar composition \citep{grev}. We computed the total
transmission coefficients as implemented in the software package PINTofALE
\citep{pintofale} from the single parameter N$_{\rm H}$. Using this model we
account only for absorption from the neutral component of the absorbing
column while absorption by ionized elements is included only for helium.

\begin{figure*}[!ht]
\resizebox{\hsize}{!}{\includegraphics{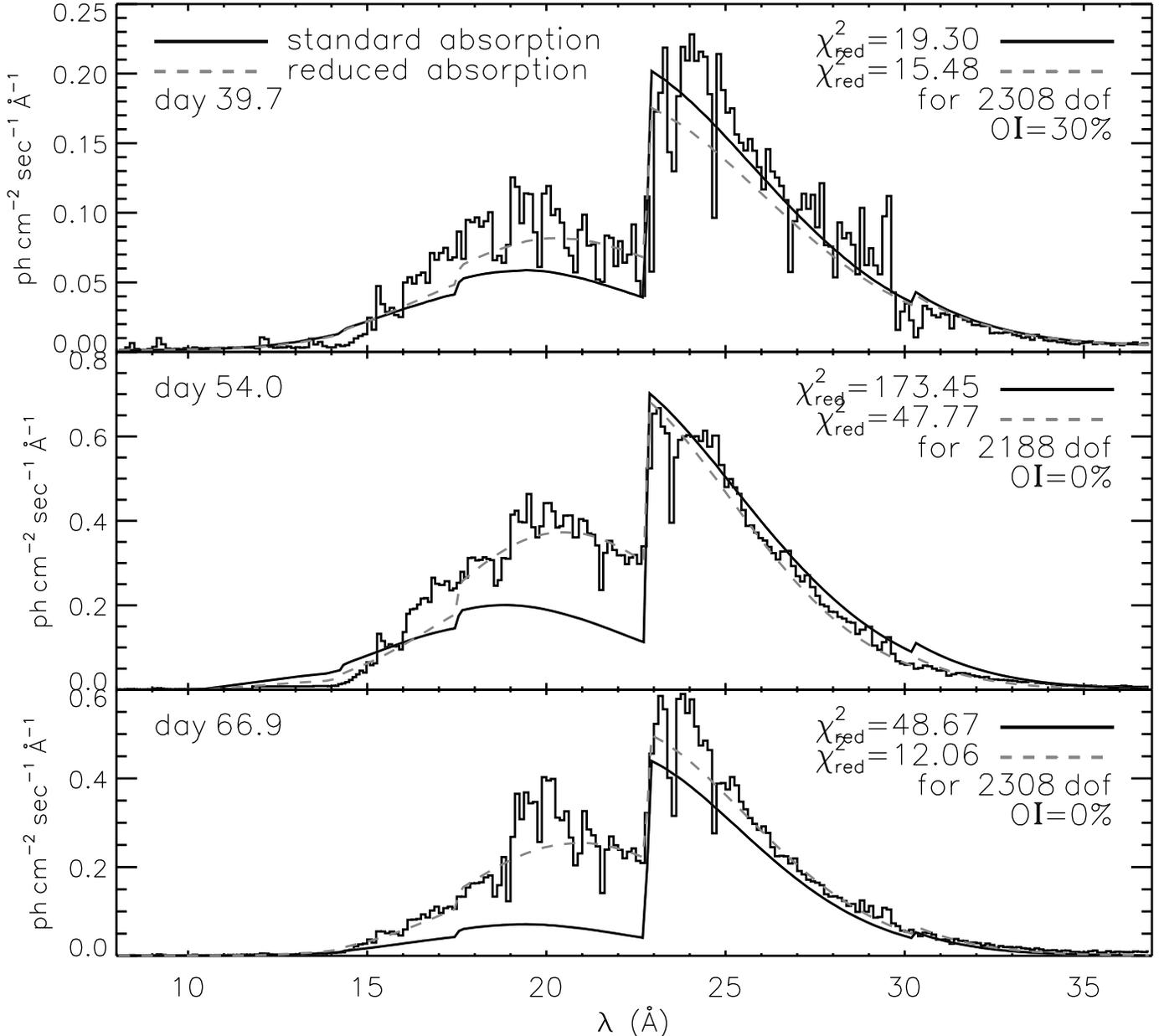}}
\caption{\label{bbb}Three grating spectra of RS\,Oph in
chronological order from top to bottom with a continuum
model and two different absorption models applied to each
case. The standard absorption models (black solid line) assume
solar abundances for the neutral elements in the line of
sight. For the second models (grey dashed) we allowed the abundance
of neutral oxygen to vary. For all observations, both models
have large $\chi^2_{\rm red}$, however, the assumption of
reduced neutral oxygen leads to an improvement of the
model.
}
\end{figure*}

In Fig.~\ref{bbb} we present the three spectra from the
gratings (in photon flux units) in comparison with our best-fit
continuum plus absorption models. First, we optimized the
blackbody parameters and the value of N$_{\rm H}$ that
represents the neutral component of the circumstellar
material with the abundances
fixed at solar \citep{grev}. In the upper right corner
of each panel we give the value of $\chi^2_{\rm red}$.
These models are identical to the ones described above,
and are formally unacceptable. The blackbody continuum
seems to represent the rough observed spectral shape.
However, the high resolution of the
gratings clearly exposes the shortcomings of blackbody
models with many spectral features not being reproduced.

 Next, we varied the abundance of the oxygen content within
the circumstellar component, while the interstellar oxygen
abundance remained fixed at solar. We fit the oxygen
abundance simultaneously with the blackbody parameters and
N$_{\rm H}$ and include the best-fit models with grey lines in
Fig.~\ref{bbb}. Although, these models are also formally
unacceptable, an improvement in $\chi^2_{\rm red}$ is
apparent. The resulting best-fit abundances suggest that
after day 54 all neutral oxygen in the circumstellar material
has disappeared. We note that for day 66.9 the optical depth
around the edge was higher than on day 54.0, suggesting that the
neutral oxygen abundance has increased again. The model for
day 66.9 does not show this increase, but the peak is not
well modeled. The blackbody temperatures of these models are
$(540-760)\times10^3$\,K on day 39.7, 
$(520-580)\times10^3$\,K on day 54.0, and $(440-590)\times10^3$\,K
on day 66.9. The bolometric luminosities, $\log(L_{\rm bol})$, of
these models are between 37.6 and 39.2 for day 39.7, 39.3 and 40.0
for day 54.0, and 38.9 and 40.5 for day 66.9. Comparison with the
parameters of the models with the oxygen abundance fixed at solar
shows that the introduction of the oxygen abundance as a free
parameter introduces considerable additional uncertainty in the
blackbody parameters.

 The simplest explanation for the reduction of neutral
oxygen is photoionization of the circumstellar material
by the radiation field. We tested this hypothesis and computed
models with reductions of other elements that would also
have to be ionized (nitrogen and carbon with their
K-shell absorption edges at 30.25\,\AA\ and 43.63\,\AA,
respectively). We were not able to find a model that
reproduced the measured spectrum better than the
one with only oxygen reduced. Next, we carefully inspected the
grating spectra for evidence of any absorption edges at the
ionization energies of O\,{\sc vii} (16.77\,\AA),
O\,{\sc viii} (14.23\,\AA), N\,{\sc vi} (22.46\,\AA), and
N\,{\sc vii} (18.59\,\AA). In all three spectra, there is no
evidence for these absorption edges. Since the
continuum level at these wavelengths is sufficiently high, and
the instruments are sensitive enough to detect these edges,
we conclude that no absorption edges from H-like and He-like
ions are present. This indicates that the material that
produces the absorption lines from high ionization stages
(see next section) contributes little bound-free
absorption and resides deeper within the outflow.

\cite{paerels01_abs} pointed out that the detailed structure
of the spectrum around the O edge is extremely complex,
consisting of absorption lines and -edges for various ions
and molecular species. For example, we studied the O\,{\sc i}
$1s-2p$ absorption line at 23.62\,\AA. On days 39.7 and 66.9
we found this line at the same wavelength of 23.51\,\AA\ and
with the same optical depth at line center of $\tau_c=1.4$.
This similarity suggests that this line was formed in a
non-changing medium (interstellar medium). The shift of this
line is similar to shifts 30-50\,m\AA\ found by
\cite{juett04} for various features from singly and doubly
ionized oxygen
in HETG spectra of seven X-ray binaries which are consistent
with discrepancies between theoretical calculations and
laboratory measurements. It is thus likely that the
23.51-\AA\ line is the O\,{\sc i} $1s-2p$ line at the
rest wavelength. We also found evidence for the $1s-2p$
absorption lines of O\,{\sc ii} to O\,{\sc v}, expected
at 23.3\,\AA, 23.11\,\AA, 22.78\,\AA, and 22.33\,\AA,
respectively (see also Table~\ref{alines}), which indicates
that we are not only dealing with atomic oxygen.

 \cite{juett04} have developed detailed models for the
wavelength range around the oxygen K-shell absorption edge.
These models allow determinations of the ratios of ionization
stages and oxygen column densities and lead to a more accurate
bound-free absorption model needed for further analyses.
Similar analyses have been carried out by \cite{page03},
however, in no other source is the oxygen edge as deep as
in RS\,Oph.

Alternatively to ionization, \cite{paerels01_abs} propose
that grain formation could reduce the O edge. In the
event of dust formation in the outer regions
of RS\,Oph, oxygen could have been locked up in grains.

\subsection{Modeling of resonance absorption and emission lines}
\label{pcygsect}

\begin{figure*}[!ht]
\resizebox{\hsize}{!}{\includegraphics{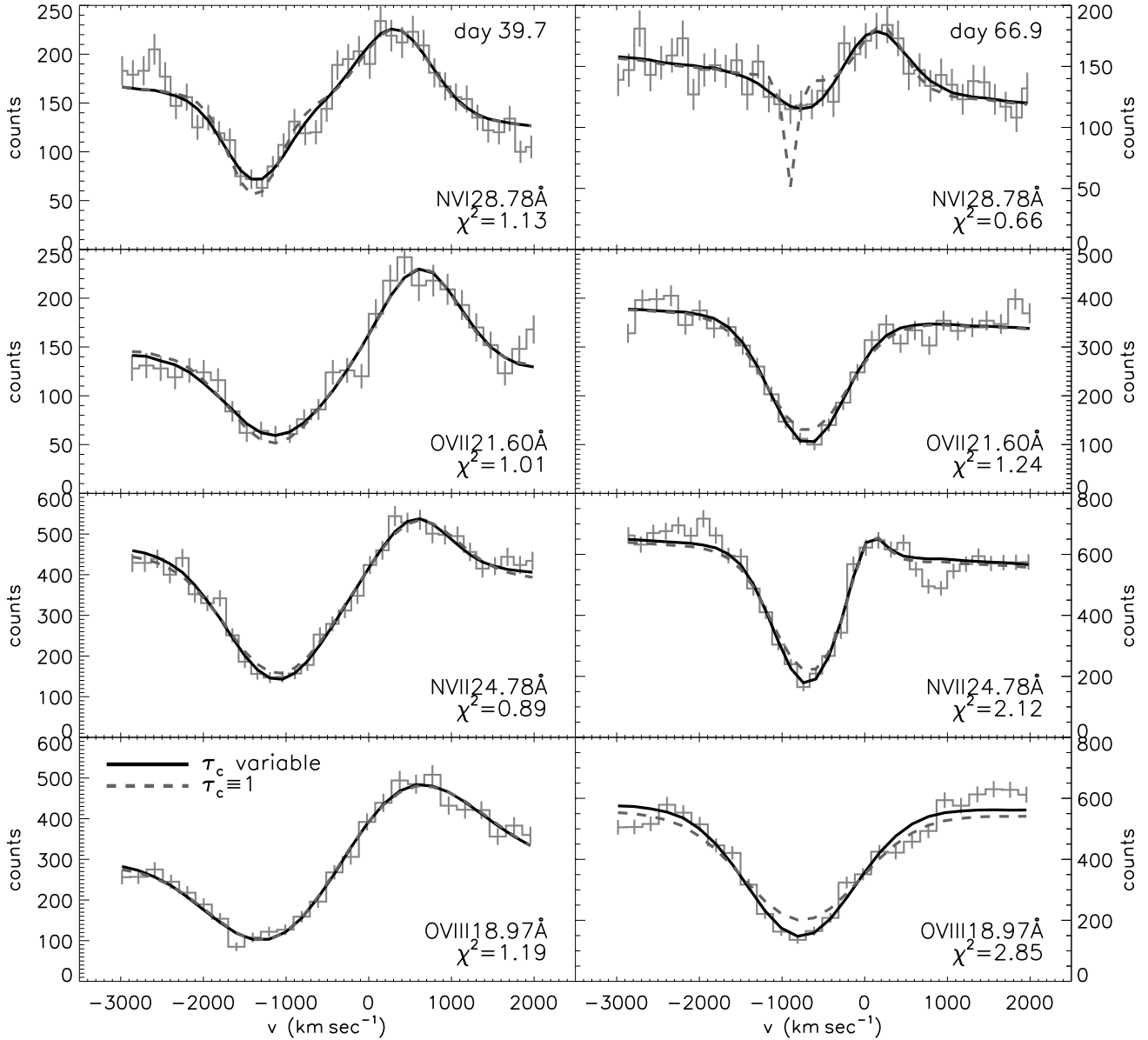}}
\caption{\label{profiles}Modeling of resonance absorption- and
emission lines detected on days 39.7 (left panels) and 66.9
(right panels) for four prominent transitions (labeled in bottom
right). The models were calculated as described in Sect.~\ref{pcygsect}
and the model parameters are shown in Table~\ref{tab_profiles}.
The dashed lines mark models where the optical depth at line
center, $\tau_c$, was fixed at unity.}
\end{figure*}

 The \chandra\ and \xmm\ grating spectra are the only spectra
that allow us to identify and study absorption- and emission
lines.
 In Fig.~\ref{profiles} we show small portions of the \chandra\
count spectra of days 39.7 and 66.9, focused on the wavelength
ranges of four prominent lines, projected on a velocity scale
(assuming the rest wavelengths $\lambda_0$
given in the legends). We modeled these spectral regions with three
components, a continuum (rescaled blackbody), an absorption line
component, and an emission line component. We include the emission
line component in order to determine if we are seeing
P-Cygni profiles \citep{rsoph_iau2}.
We restrict the model to the narrow spectral region around each line
\begin{eqnarray}
\label{pcygmodel}
M(\lambda)&=&[ \{C\times B_\lambda(T)+E\times G_e(\lambda,\lambda_e,\sigma_e)\}\\ \nonumber
          & &\ \times \{1-A\times G_a(\lambda,\lambda_a,\sigma_a)\} ]\\
          & & \times T_{\rm ISM}(\lambda)\times T_{\rm CS}(\lambda) \nonumber
\end{eqnarray}
with $C\times B_\lambda(T)$ being the continuum (blackbody model) and
$A\times G_a(\lambda,\lambda_a,\sigma_a)$ and $E\times
G_e(\lambda,\lambda_e,\sigma_e)$ being the absorption- and emission line
components, respectively, both modeled as Gaussians. In an iterative process
we optimized the normalization $C$ of the blackbody continuum at the fixed
temperature $T=6.2\times 10^5\,{\rm K}$ (based on the blackbody parameters
found in Sect.~\ref{continuum}) and the central wavelengths
($\lambda_a$ and $\lambda_e$), line widths ($\sigma_a$ and $\sigma_e$),
and normalizations ($A$ and $E$) of the absorption and emission line
components, respectively. Finally, we corrected the model for interstellar
and circumstellar absorption using the same absorption model described
in Sect.~\ref{continuum} (without the reduction of neutral oxygen)
with $T_{\rm ISM}(\lambda)$ and $T_{\rm CS}(\lambda)$ the transmission
coefficients for interstellar and circumstellar material
with N$_{\rm H}({\rm ISM})=2.4\times10^{21}$\,cm$^{-2}$ and
N$_{\rm H}({\rm CS})=2.3\times10^{21}$\,cm$^{-2}$, respectively
(equivalent to an effective H\,{\sc i} column density of
$4.7\times10^{21}$\,cm$^{-2}$).

For a given transition with the rest wavelength $\lambda_0$, the parameters
$\lambda_a$ and $\lambda_e$ can be converted to velocities from line shifts
($v_s$) for the absorption- and emission line components, respectively.
This velocity represents the expansion velocity at the radial distance
within the outflow where the optical depth is unity.
The line widths $\sigma_a$ and $\sigma_e$ are
representative of the velocity distribution of the observed
regions which depend on the velocity- and density profile within the
outflow, however, we have not accounted for the contributions of the
instrument profile and can therefore not use these numbers. We derive
emission line fluxes from the parameter $E$ and optical depths at line
center,
$\tau_c=\ln \{M(\lambda_{\rm min})/{\rm CONT}(\lambda_{\rm min})\}$ at
the wavelength where the model finds its minimum, $\lambda_{\rm min}$.
${\rm CONT}(\lambda_{\rm min})=C\times B(\lambda_{\rm min},T)
\times T_{\rm ISM}(\lambda_{\rm min})\times T_{\rm CS}(\lambda_{\rm min})$
is the continuum flux at $\lambda_{\rm min}$. Note, that the emission
line fluxes represent the intrinsic line fluxes, including
self-absorption, corrected for interstellar plus circumstellar
absorption. Integration over the entire absorption line profile
allows us to estimate absorbing column densities, $xn_e$ for each line.
We use the oscillator strengths given in Table~\ref{tab_profiles}.
The emission line fluxes and column densities depend strongly on the
choice of N$_{\rm H}({\rm ISM})+$N$_{\rm H}({\rm CS}$), especially
at long wavelengths.

We calculated uncertainty ranges for the model parameters $\lambda_{a,e}$,
$\sigma_{a,e}$, $A$, and $E$ separately by
calculating a grid of $\chi^2_{\rm red}$ over a given range of
the respective parameters. For each grid point all other
parameters were optimized, respecting the boundary condition of the
individual parameter of interest to be fixed at
the given grid value. The 1-$\sigma$ uncertainty ranges were then obtained
by interpolating the grid until $\chi^2_{\rm red}$ had increased by unity.
If the curve was found to be too flat so that $\chi^2_{\rm red}$ never reached
values higher than one above the minimum for a sensible range of values,
then lower or upper limits were calculated. Next, the uncertainties of
$\lambda_{a,e}$ and $\sigma_{a,e}$ were translated into the respective
uncertainties in velocity, and those of $A$ and $E$ into the uncertainties
of $\tau_c$ and emission line fluxes, respectively. For $\tau_c$ we only
considered variations of $A$, but we allowed all other parameters to be
optimized before a value of $\chi^2_{\rm red}$ was calculated for a given
$\tau_c$. The uncertainties of the column densities
were obtained by calculating a grid of values of $\chi^2_{\rm red}$ from
a range of pairs of $A$ and $\sigma_a$ that represent a grid of
values of $xn_0$ and probing the range of $\chi^2_{\rm red}+1$.

 The uncertainties of $\tau_c$ and $xn_0$ do not explore the full
uncertainty range as variations of
the central wavelength of the emission line component can dramatically
change those values. In particular if saturation is reached, we loose
all constraints. This is a shortcoming of our parameterized model that
cannot deal with saturation self-consistently.
For now, the given uncertainty ranges represent the minimal
uncertainties within which no changes can be considered significant.
We have also not included uncertainties from the amount of interstellar
and circumstellar absorption.

 In Fig.~\ref{profiles} we compare the count spectra with the models
$M(\lambda)$
after iteration of $C$, $E$, $A$, $\lambda_e$, $\sigma_e$, $\lambda_a$,
and $\sigma_a$ (Eq.~\ref{pcygmodel}). The relevant
spectral ranges are well fitted, and the values of $\chi^2_{\rm red}$
are below 1.2 for day 39.7 and below 3.0 for day 66.9.
 In Table~\ref{tab_profiles} we list the line-shift
velocities, optical depths at line center, and emission line fluxes,
derived from the model parameters with their uncertainty ranges for
the absorption- and emission line components.

The optical depths in the line centers are unity within the uncertainties
for all lines in both observations, which is consistent with our
expectation that we are observing regions in the outflow
where $\tau=1$. In Fig.~\ref{profiles} we add a dashed lines denoting
the respective models where $\tau=1$ was enforced during the
fit. For N\,{\sc vi} (day 66.9) we were not able to constrain any of
the parameters because the absorption trough is quite flat. An optical
depth of $\tau=1$ leads only to an acceptable fit
if the absorption line is assumed to be very narrow and
leads to an increase in $\chi^2$ in only one spectral bin (see top
right panel of Fig.~\ref{profiles}).

The column densities could not be sufficiently constrained to assess
whether they varied significantly from day 39.7 to 66.9. Only
N\,{\sc vii}, and possibly N\,{\sc vi}, exhibited a decrease of column
density within the given errors. However, we
caution that uncertainties from the emission line component and the
continuum as well the amount of interstellar+circumstellar absorption
are not included in these error estimates.

Within each observation the absorption line velocities are
consistent with \vaverageone\ for day 39.7 and \vaveragetwo\ for
day 66.9 (weighted averages). This marks a clear reduction in the
measured expansion velocity. The line widths, although dominated by
the instrumental line profile, have systematically but not significantly
decreased. We tested models where we fixed all line widths at the
instrumental resolution ($\sigma_{\rm gauss}=0.03$\,\AA\ at all
wavelengths) and found acceptable fits except for N\,{\sc vii}.

 The potential of this
approach over global approaches is that we can concentrate on single
lines that are least affected by blending or other uncertainties that
are more difficult to disentangle in complex atmosphere models. We
found optical depths at line centers of unity.
The second reliable result at this stage is the decrease of the
characteristic velocity consistently measured for five different
absorption lines. 
Although the velocities derived from these absorption lines are very 
similar to those expected at these epochs from the simple shock model as 
applied to the Swift X-ray data \citep{bode06}, similar effects in 
these lines observed in X-rays in classical novae are commonly ascribed
to observing material deeper into the envelope in which there is a
velocity gradient (Hubble flow). Further, the origin of the emission
line components has to be determined before further steps can be
taken. This requires a reduction of the uncertainties which we plan
to achieve by applying sensible boundary conditions.

\begin{table*}[!ht]
\begin{flushleft}
\renewcommand{\arraystretch}{1.1}
\caption{\label{tab_profiles}Results from profile fitting illustrated in
Fig.~\ref{profiles}}
{\small
\begin{tabular}{p{.1cm}rr|rr|rr|rr|rr}
\hline
ion                & \multicolumn{2}{c}{O\,{\sc vii}}   & \multicolumn{2}{c}{O\,{\sc viii}}
                   & \multicolumn{2}{c}{N\,{\sc vi}}    & \multicolumn{2}{c}{N\,{\sc vii}}
                        & \multicolumn{2}{c}{Fe\,{\sc xvii}}\\
$\lambda_0$        & \multicolumn{2}{c}{21.60\,\AA}     & \multicolumn{2}{c}{18.97\,\AA}
                   & \multicolumn{2}{c}{28.78\,\AA}     & \multicolumn{2}{c}{24.78\,\AA}
                        & \multicolumn{2}{c}{15.01\,\AA}\\
$f^{[a]}$		   & \multicolumn{2}{c}{0.68}     & \multicolumn{2}{c}{0.83}
                   & \multicolumn{2}{c}{0.66}     & \multicolumn{2}{c}{0.83}
                        & \multicolumn{2}{c}{2.52}\\

    &  day 39.7  & day 66.9   & day 39.7   & day 66.9   & day 39.7    & day 66.9  & day 39.7  & day 66.9& day 39.7  & day 66.9\\
\multicolumn{9}{l}{\bf Absorption Component}\\
$^{[b]}v_{\rm s}$      & $-1143^{+939}_{-317}$ & $-685^{+101}_{-104}$     & $-1264^{+477}_{-254}$ & $-772^{+91}_{-87}$
                   & $-1365^{+907}_{-227}$ & --			      & $-1098^{+444}_{-116}$ & $-701^{+407}_{-78}$
                   & $-1556^{+661}_{-426}$ & $-834^{+439}_{-428}$\\

%
$^{[c]}\tau_c$         & $0.79^{+0.83}_{-0.46}$ & $1.23^{+0.94}_{-0.45}$  & $1.04^{+0.46}_{-0.85}$ & $1.37^{+0.92}_{-0.44}$
                   & $0.72^{+0.97}_{-0.40}$ & $<3.0$                  & $1.08^{+0.45}_{-0.40}$ & $1.01^{+0.49}_{-0.24}$
                   & $0.87^{+1.93}_{-0.87}$ & $0.79^{+1.81}_{-0.60}$\\

$^{[d]}xn_0$           & $3.38^{+1.04}_{-0.96}$ & $3.17^{+0.61}_{-0.43}$  & $4.53^{+0.59}_{-0.69}$ & $4.53^{+0.44}_{-0.50}$
                   & $1.88^{+0.47}_{-0.54}$ & $<1.3$			  & $3.25^{+0.35}_{-0.40}$ & $2.42^{+0.24}_{-0.30}$
                   & $1.23^{+0.81}_{-0.66}$ & $1.25^{+0.54}_{-0.57}$\\
\multicolumn{9}{l}{\bf Emission Component}\\
$^{[b]}v_{\rm s}$     & $633^{+316}_{-371}$ & -- & $544^{+402}_{-1334}$ & --
                  & $329^{+250}_{-1839}$ & --			        & $559^{+593}_{-735}$ & $3^{+583}_{-335}$
                  & $363^{+383}_{-1969}$ & --\\

$^{[e]}$flux          & $3.78^{+3.5}_{-2.0}$  & -- & $2.20^{+3.0}_{-0.9}$ & --
                  & $31.9^{+18}_{-15}$ &   --                          & $3.55^{+7.52}_{-2.56}$ & $2.87^{+14}_{-2.9}$
                  & $0.13^{+0.13}_{-0.05}$ & --\\
\hline
\end{tabular}
}

{\scriptsize
$^{[a]}$oscillator strength \ \ \ \ \ \ $^{[b]}$ velocity from line shift (km\,s$^{-1}$)\hfill
 $^{[c]}$optical depth at line center\\
$^{[d]}$column density (10$^{16}$cm$^{-2}$)\hfill
$^{[e]}$emission line flux ($10^{-10}$erg\,cm$^{-2}$sec$^{-1}$)\\
$^{[d]}$uncertainties of $xn_0$ were calculated at fixed wavelengths and 
 with fixed emission line parameters and are underestimated\\
}
\renewcommand{\arraystretch}{1}
\end{flushleft}
\end{table*}

 For the observation taken on day 39.7, we first assumed only absorption 
lines, but an additional emission line component was necessary in
order to arrive at acceptable fits. Their nominal (line-shift)
velocities range from $\sim +300$ to $+600$\,km\,s$^{-1}$, but the
uncertainty ranges are large and include the possibility that the
emission lines could be at their rest wavelengths.

We scanned the \chandra\ spectrum of day 39.7 for all absorption lines
that appear not to be blended and summarize the wavelengths and central
optical depths $\tau_c$ below the continuum in Table~\ref{alines}. We
attempted identifications and found lines from N, O, Fe, S, Si, Ca, and
Ar using the collisional database APED\footnote{Astrophysical Plasma
Emission Database v1.3 \citep{smith01}}. The identifications are based
on the assumption that a strong collisional excitation probability
implies a strong radiative excitation probability. We searched for
candidates at wavelengths that imply blue shifts of
$\sim 0.1$\,\AA, as this wavelength shift is found for
the well-identified lines used for Fig.~\ref{profiles}.

\begin{table}[!ht]
\begin{flushleft}
\renewcommand{\arraystretch}{1.1}
\caption{\label{alines}Strongest absorption lines found in ObsID 7296 (day 39.7)}
\begin{tabular}{cccr||cccr}
$\lambda^{[a]}$ & $\tau_c$ & $\lambda_0^{[a]}$ & ID$^{[b]}$ & $\lambda^{[a]}$ & $\tau_c$ & $\lambda_0^{[a]}$ & ID$^{[b]}$\\
\hline
14.93  &  0.86  	   & 15.01 & Fe\,{\sc xvii}   & 23.51  &  1.41  & 23.62 & O\,{\sc i}\\
15.77  &  0.41  	   & 15.87 & Fe\,{\sc xviii}? & 23.66  &  0.35  & 23.78 & N\,{\sc vi}\\
15.94  &  0.92  	   & 16.00 & O\,{\sc viii}    & 24.50  &  0.24  & 24.51 & S\,{\sc xiv}?\\
16.71  &  0.43  	   & 16.78 & Fe\,{\sc xvii}   & 24.69  &  0.93  & 24.78 & N\,{\sc vii}\\
16.99  &  0.51  	   & 17.05 & Fe\,{\sc xvii}   & 25.61  &  0.57  & 25.68 & Ar\,{\sc xvi}?\\
17.69  &  0.48  	   & 17.77 & O\,{\sc vii}?    & 25.76  &  0.25  & 25.84 & Ar\,{\sc xv}?\\
18.88  &  1.00  	   & 18.97 & O\,{\sc viii}    & 25.91  &  0.21  & 26.00 & Ar\,{\sc xv}?\\
19.84  &  0.75		   & 19.83 & N\,{\sc vii}     & 26.07  &  0.51  & 26.12 & N\,{\sc vi} \\
20.63  &  0.43  	   & 20.68 & S\,{\sc xiv}?    & 26.82  &  1.06  & 26.99 & C\,{\sc vi}?\\
20.84  &  0.44  	   & 20.91 & N\,{\sc vii}     & 27.30  &  0.42  & 27.47 & Ar\,{\sc xiv}?\\
21.52  &  0.83  	   & 21.60 & O\,{\sc vii}     & 27.52  &  0.79  & 27.64 & Ar\,{\sc xiv}?\\
22.25  &  0.56		   & 22.33 & O\,{\sc v}       & 27.83  &  0.49  & 27.90 & Si\,{\sc xii}?\\
22.64  &  1.18  	   & 22.78 & O\,{\sc iv}      & 28.00  &  0.55  & 28.11 & Fe\,{\sc xx}?\\
22.93  &  1.06  	   & 22.97 & Ca\,{\sc xvi}?   & 28.65  &  0.70  & 28.78 & N\,{\sc vi}\\
23.16  &  0.31  	   & 23.11 & O\,{\sc iii}     & 29.80  &  0.66  & 29.91 & Fe\,{\sc xvii}?\\
\multicolumn{2}{r}{\bf or} & 23.29 & N\,{\sc vi}      & 30.35  &  1.59  & 30.47 & Ca\,{\sc xi}?\\
23.36  &  0.25		   & 23.30 & O\,{\sc ii}      & 31.29  &  0.94  & 31.32 & Fe\,{\sc xviii}?\\
\hline 
\end{tabular}

$^{[a]}$in \AA\hfill $^{[b]}$Uncertain identifications are marked with '?'
\renewcommand{\arraystretch}{1}
\end{flushleft}
\end{table}

\subsection{Emission lines from the shock}
\label{elinesect}

\begin{figure}[!ht]
\resizebox{\hsize}{!}{\includegraphics{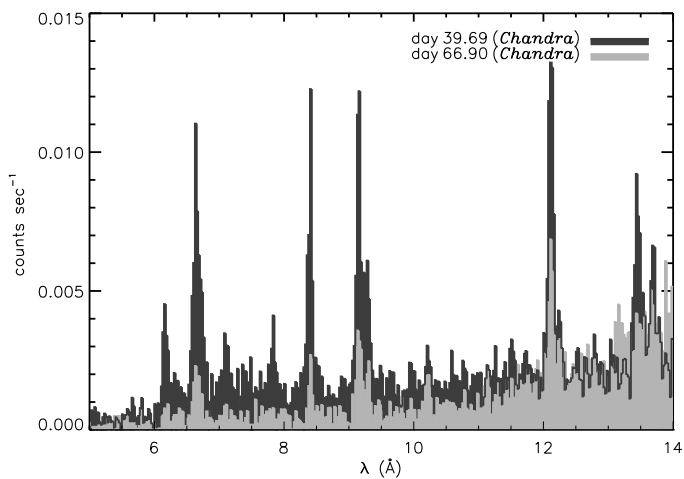}}
\caption{\label{cmp_sss_lines}Comparison of the \chandra\ spectra taken
on days 39.7 (dark) and 66.9 (light) focusing on the high-energy emission
lines. These lines are independent of the SSS spectrum and are fading.}
\end{figure}

Shortwards of 14\,\AA, we found emission lines at wavelengths where
\chandra\ and \xmm\ measured shock-induced emission lines before
the SSS spectrum appeared
\citep[][Ness in prep.]{rsoph_iau1,2006IAUC.8695....2G}.
In Fig.~\ref{cmp_sss_lines} we show the count rate spectra of this
spectral range for the \chandra\ observations taken on days 39.7 and
66.9. In contrast to the SSS spectra on these days, the latter spectrum
exhibits fainter emission lines than during the day-39.7 observation.
These lines must be collisionally excited, due to the absence of any
ionizing radiation of sufficient energies, and we interpret them as
residual emission from the shock.

We measured wavelengths and emission line fluxes using our line
fitting tool Cora \citep{newi02}. The wavelengths can be converted
to velocities that characterize the dynamics of the plasma that
produces these lines. For day
39.7 we found a small systematic line shift from all lines that
translate to a weighted average plus variance of
$-526\,\pm\,114$\,km\,s$^{-1}$. For the other two observations,
the lines were not strong enough to estimate reliable uncertainties
of the line positions.

\begin{table*}[!ht]
\begin{flushleft}
\renewcommand{\arraystretch}{1.1}
\caption{\label{elines}Evolution of emission line fluxes.}
\begin{tabular}{lrccccc|ccc}
& & &  & flux$^{[c]}$ & flux$^{[c]}$& flux$^{[c]}$ & ratio & ratio & ratio\\
\multicolumn{2}{l}{Ion \hfill $\lambda_0^{[a]}$\small{(\AA)}}& TC$^{[b1]}$ & TC$^{[b2]}$ & day 39.6 & day 54.0 & 66.9 & 54/39 & 66/54 &66/39\\
\hline
\multicolumn{7}{l|}{\bf Lines from shock ('hard lines')}&&\\
Si\,{\sc xiv}    & 6.18  & 0.841 &  0.897 & $112.1\,\pm\,10.6$	 & $ 47.2\,\pm\,12.8$   & $22.1\,\pm\,6.3$   &  0.42 & 0.47 &  0.20 \\
Si\,{\sc xiii}   & 6.65  & 0.809 &  0.876 & $336.4\,\pm\,55.5$	 & $158.3\,\pm\,68.2$   & $50.9\,\pm\,33.1$  &  0.47 & 0.32 &  0.15 \\
Mg\,{\sc xii}    & 8.42  & 0.693 &  0.793 & $196.8\,\pm\,12.3$	 & $106.7\,\pm\,7.9 $   & $40.5\,\pm\,7.2$   &  0.54 & 0.38 &  0.21 \\
Mg\,{\sc xi}     & 9.20  & 0.626 &  0.744 & $386.3\,\pm\,38.1$	 & $240.2\,\pm\,26.5$   & $133.3\,\pm\,27.7$ &  0.62 & 0.56 &  0.35 \\
Ne\,{\sc x}      & 12.13 & 0.390 &  0.556 & $285.5\,\pm\,11.8$	 & $105.7\,\pm\,9.3 $   & $95.2\,\pm\,8.3$   &  0.37 & 0.90 &  0.33 \\
Fe\,{\sc xvii}   & 12.26 & 0.380 &  0.547 & $ 70.2\,\pm\,11.5$ 	 & $ 17.4\,\pm\,8.1 $	& $26.5\,\pm\,8.6$   &  0.25 & 1.52 &  0.38 \\
Ne\,{\sc ix}     & 13.44 & 0.290 &  0.463 & $138.1\,\pm\,13.8$	 & $ 42.0\,\pm\,5.5 $   & $34.1\,\pm\,11.4$  &  0.30 & 0.81 &  0.25 \\
Ne\,{\sc ix}$^{[i]}$ & 13.55 & 0.282 &  0.455 & $ 66.7\,\pm\,8.0 $   & $ 28.8\,\pm\,4.1 $   & $38.1\,\pm\,8.1$   &  0.43 & 1.32 &  0.57 \\
Ne\,{\sc ix}$^{[i]}$ & 13.69 & 0.273 &  0.445 & $104.4\,\pm\,9.9 $ 	 & $ 36.4\,\pm\,3.8 $   & $52.5\,\pm\,8.4$   &  0.35 & 1.45 &  0.50 \\
Fe\,{\sc xviii}  & 14.20 & 0.239 &  0.410 & $ 48.7\,\pm\,6.6 $   &	  --		& $20.7\,\pm\,9.1$   &	 &	&  0.42\\
\hline
\multicolumn{7}{l|}{\bf Lines on top of the SSS spectrum ('soft lines')}&&\\
Fe\,{\sc xvii}   & 15.01 & 0.225 &  0.403 & $170^{+170}_{-70}$     	 &	  -- 	 	  & --&&&\\
O\,{\sc viii}    & 18.97 & 0.087 &  0.226 & $848^{+1140}_{-130}$	 &	  --		  & --&&&\\
O\,{\sc vii}     & 21.60 & 0.031 &  0.121 & $361^{+340}_{-190}$ 	 &	  --		  & --&&&\\
O\,{\sc vii}$^{[i]}$ & 21.80 & 0.028 &  0.115 & $381.8\,\pm\,24.6$		 &	  --		  & --&&&\\
O\,{\sc vii}$^{[i]}$ & 22.10 & 0.025 &  0.106 & $418.2\,\pm\,24.6$		 &	  --		  & --&&&\\
N\,{\sc vii}     & 24.78 & 0.051 &  0.181 & $400^{+850}_{-290}$ 	 &	  --		  & $321^{+1380}_{-320}$&&&\\
N\,{\sc vi}      & 28.78 & 0.010 &  0.072 & $325^{+180}_{-150}$ 	 &	  --		  & $247^{+920}_{-250}$ &&&\\
N\,{\sc vi}$^{[i]}$  & 29.10 & 0.009 &  0.066 & $282.1\,\pm\,19.1$ 		 &	  --		  & $443.9\,\pm\,31.9$&&&\\
N\,{\sc vi}$^{[i]}$  & 29.54 & 0.007 &  0.059 & $788.5\,\pm\,27.3$		 &	  --		  & $397.7\,\pm\,32.6$&&&\\
\hline
\end{tabular}

$^{[a]}$rest wavelengths\ \ \ $^{[b1], [b2]}$transmission coefficients
assuming $^{[b1]}$N$_{\rm H}=4.7\times10^{21}$\,cm$^{-2}$
and $^{[b2]}2.4\times10^{21}$\,cm$^{-2}$\\
$^{[c]}10^{-14}$\,erg\,cm$^{-2}$\,s$^{-1}$, not corrected for interstellar absorption\ \
$^{[i]}$intersystem line
\renewcommand{\arraystretch}{1}
\end{flushleft}
\end{table*}

In Table~\ref{elines} we list the emission line fluxes for the strongest
lines. We have not corrected the measured fluxes for the effects of
interstellar and/or circumstellar absorption. As shown in
Sect.~\ref{continuum}, our models of interstellar and circumstellar
absorption are not accurate enough to derive reliable corrections. In
the first four columns we list the line identifications, wavelengths,
and transmission coefficients that would have to be applied under the
assumptions of
N$_{\rm H}({\rm ISM})+$N$_{\rm H}({\rm CS})=4.7\times10^{21}$\,cm$^{-2}$
and N$_{\rm H}({\rm CS})=0$ (i.e., only interstellar absorption with
N$_{\rm H}({\rm ISM})=2.4\times10^{21}$\,cm$^{-2}$) assuming
solar abundances. As soon as a more accurate absorption model is
available, the fluxes listed in Table~\ref{elines} can be converted
to unabsorbed fluxes at the source. However,
if the lines originate from the shock, they may only be affected by
interstellar absorption and not by circumstellar material.
In the top part of Table~\ref{elines} we list the line fluxes
originating from the shock and in the bottom part we include the
emission line fluxes measured as part of the models described in
Sect.~\ref{pcygsect} for days 39.7, 54.0, and 66.9. We reference we
label these lines 'hard lines' and 'soft lines', respectively.
For the soft lines we reversed the correction for interstellar
absorption that has been applied with the terms
$T_{\rm ISM}(\lambda)\times T_{\rm CS}(\lambda)$ in Eq.~\ref{pcygmodel}
for comparison with the unabsorbed hard lines. We also include the line
fluxes for He-like intersystem lines measured in Sect.~\ref{hesect}.
In the last three columns we list the changes in the line fluxes
between days 54.0 and 39.9, 66.9 and 54.0, and 66.9 and 39.7,
respectively.

The fluxes are all of the same order of magnitude which implies that
they originate from the same plasma. In that case the soft
lines considered as part of the absorption lines in
Sect.~\ref{pcygsect} are actually coming from the
shock. This would also explain why these lines disappeared by day 66.9,
which earlier had been interpreted as disappearance of P Cygni profiles
\citep{rsoph_iau2}. However, correcting for interstellar plus
circumstellar absorption will result in considerable enhancements
of the fluxes of the soft lines because of their longer wavelengths,
but that depends critically on the value of
N$_{\rm H}$(ISM)$+$N$_{\rm H}$(CS) and the applied absorption model.
The resulting higher fluxes in the soft lines
can be explained by photoexcitation in addition
to collisional excitations. If the lines originate from the shock,
only interstellar absorption has to be accounted for, leading to line
fluxes that suggest a mixture of collisional
excitations and photoexcitations. Otherwise, circumstellar absorption
has to be added, and stronger line fluxes would have to be assumed
that could be explained by a smaller distance between the
radiation source and the plasma emitting the soft lines.

\subsection{He-like triplets}
\label{hesect}

 The most prominent transitions in the range 5\,\AA --45\,\AA\ are H-like
and He-like lines of elements with nuclear charge $Z=6$ (carbon) to $Z=14$
(silicon). The $n=2-1$ transitions of He-like ions consist of three strong
lines, a resonance line (r: 1s2p\,$^1$P$_1$) an intercombination line
(i: 1s2p\,$^3$P$_1$), and a forbidden line (f: 1s2s\,$^3$S$_1$), all
decaying into the ground state \citep[1s$^2\,^1$S$_0$, see,
e.g.,][]{ness_CS}. These lines are important for two reasons. First,
the measured wavelengths of the intercombination- and forbidden lines
can constrain the exact location of the emission line components in
the models described in Sect.~\ref{pcygsect} because they are not
absorbed. Second, the flux ratio f/i of the He-like triplets are
important density diagnostics \citep[e.g.,][]{gj69,denspaper}. In
low-density environments with the presence of strong UV fields, the
f/i ratios can also be used to measure the distance between the X-ray
plasma and the UV source \citep[applied, e.g., by][to the winds of
O stars immersed
in the UV radiation field from the stellar surface]{waldcas01}.

 We used our line fitting tool Cora \citep{newi02} to determine the
line positions and line fluxes of the He-like lines of O\,{\sc vii},
$\lambda_0=$(21.6, 21.8, 22.1)\,\AA, and N\,{\sc vi}
$\lambda_0=$(28.78, 29.1, 29.54)\,\AA,
using Gaussian line profiles. We varied the line
positions and found that the i and f lines are not significantly
shifted from their rest wavelengths (upper limit
$\sim 600$\,\,km\,s$^{-1}$). More refined models in
Sect.~\ref{pcygsect} can therefore be calculated with the boundary
condition that the emission line components are not shifted.

\begin{figure}[!ht]
\resizebox{\hsize}{!}{\includegraphics{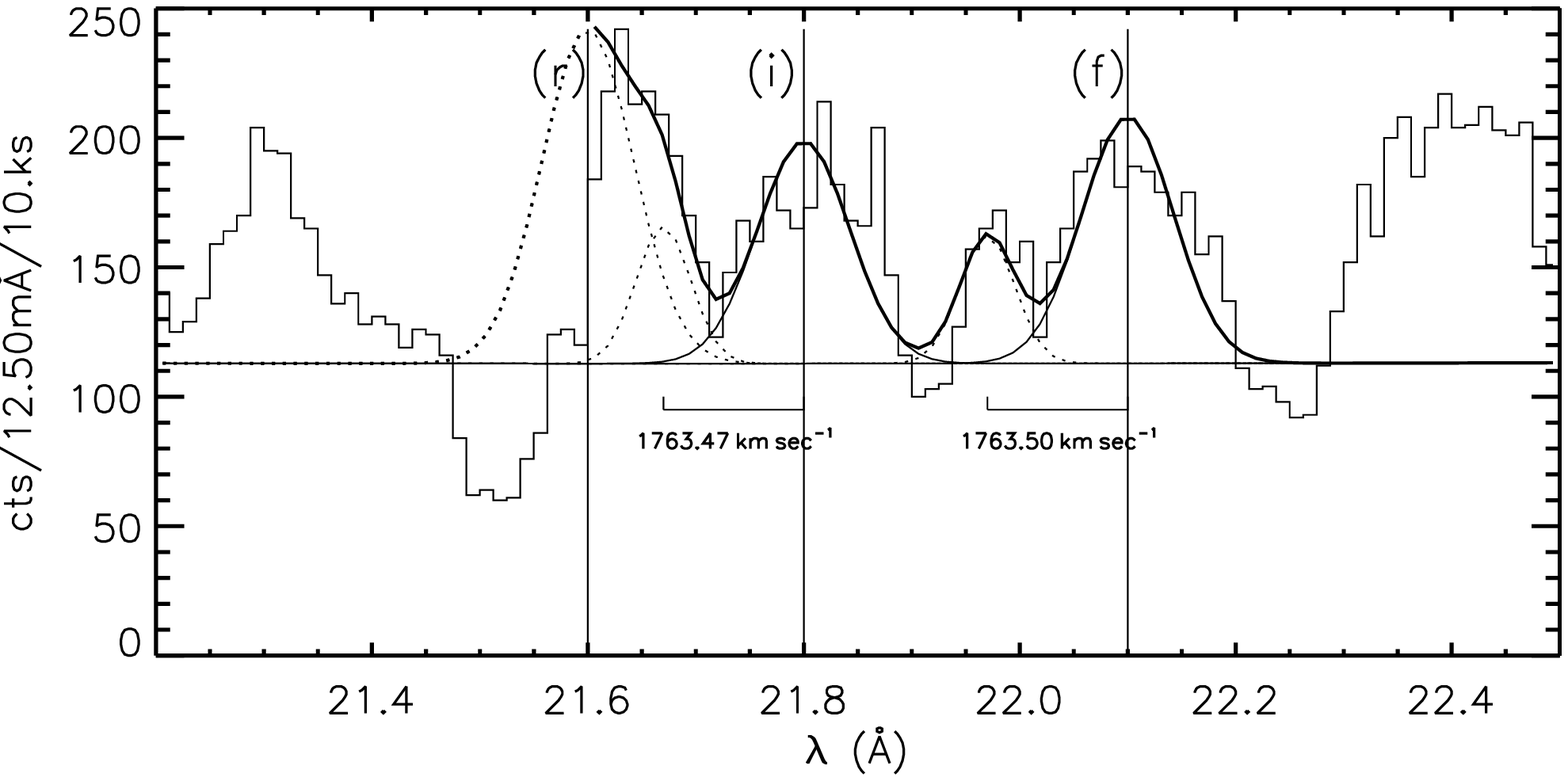}}

\resizebox{\hsize}{!}{\includegraphics{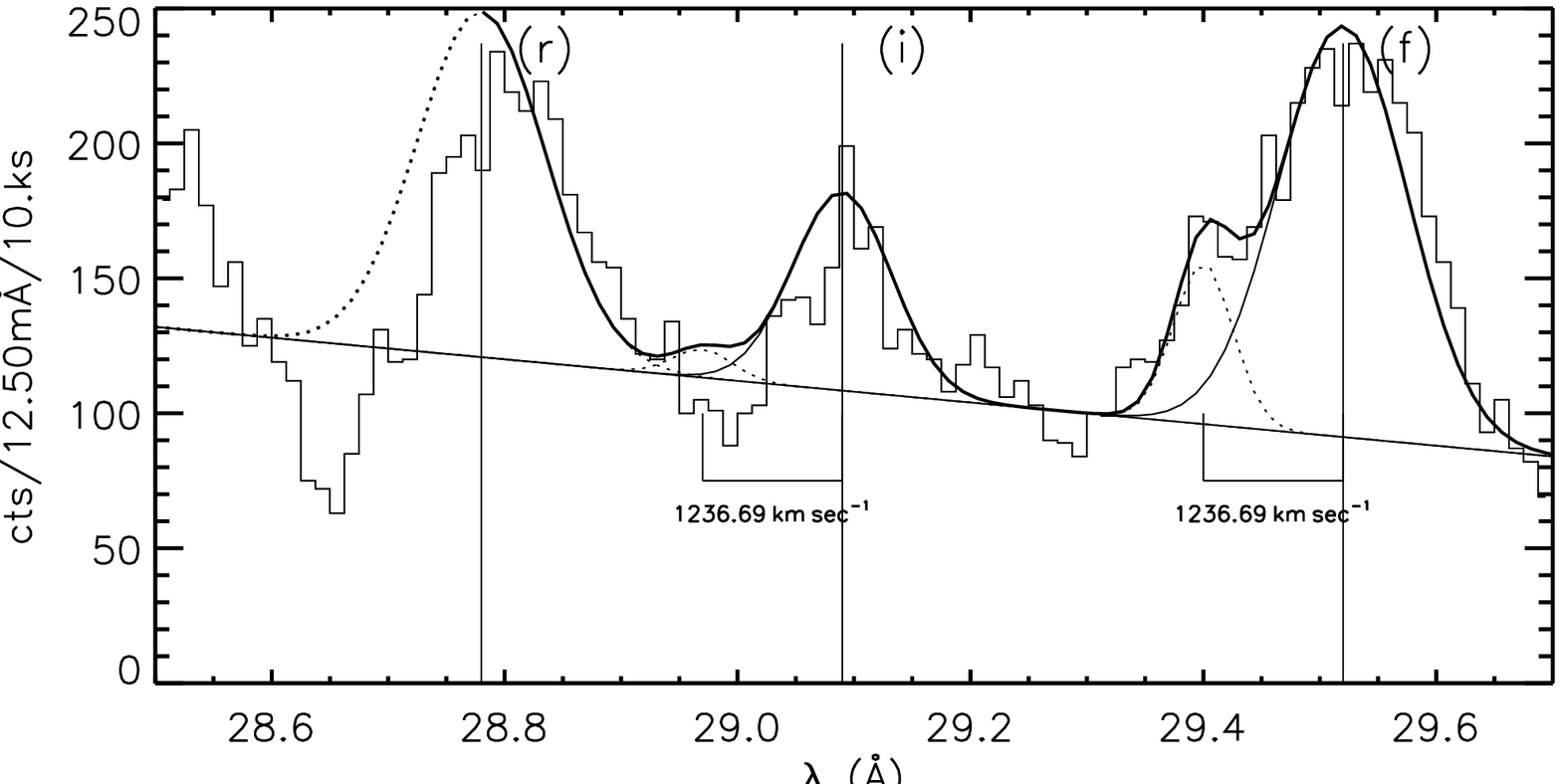}}
\caption{\label{o7}Spectral region around the He-like triplets of
O\,{\sc vii} (top) and N\,{\sc vi} (bottom) on day 39.7. The respective
resonance lines (r) at 21.6 and 28.78\,\AA\ are seen in absorption and
emission while, the intercombination- (i) and forbidden (f) lines
are only seen in emission, giving us the likely wavelengths of
all emission lines. Gaussian fits to the lines
reveal no line shifts (upper limit $\sim 600$\,km\,s$^{-1}$).
Additional lines are seen shortwards of the forbidden lines,
that could come from a velocity component with $-1763$\,km\,s$^{-1}$
and $-1237$\,km\,s$^{-1}$ for O\,{\sc vii} and N\,{\sc vi},
respectively, but no counterpart in the N\,{\sc vi} intercombination
line is detected.}
\end{figure}

 Next, we measured the line fluxes with variable line strengths and
fixed wavelengths and line widths on top of a constant continuum.
For O\,{\sc vii} we used a constant value of 9000\,counts per \AA\
and for N\,{\sc vi} where the continuum is not as flat as for
O\,{\sc vii}, we used a linear continuum with a slope of -40
counts per \AA. In Fig.~\ref{o7} we show the spectra and
our models for day 39.7.

 We considered the possibility that the lines shortwards of the
forbidden lines ($\sim 22$\,\AA\ and $\sim 29.4$\,\AA) are
blue-shifted counterparts of the forbidden lines. In that case
the intercombination lines should also have blue-shifted companions
at the same velocity. For O\,{\sc vii} a model with an additional
component,
shifted by $-1763.5$\,km\,s$^{-1}$, results in a good representation of
the data (top panel of Fig.~\ref{o7}). For N\,{\sc vi} the forbidden
line clearly shows a second component at a different velocity, but there
is no evidence for
a blue-shifted component of the intercombination line at this
velocity. No transitions from other ions are known for these
wavelengths that would be strong enough, but these lines could be
satellite lines that appear in a photoionized recombining plasma.
For our further analysis we only use the fluxes measured at the rest
wavelengths.

 We include the measured fluxes for the i- and f lines in
Table~\ref{elines}. In a collisional plasma the sum of the fluxes
in the i- and f- lines is expected to be of the same order
as that in the resonance line (r), while in a plasma where
photoexcitation takes place, the ratio $G=$(f+i)/r should be less
than 1. If this ratio is greater than one, this indicates a purely
recombining plasma or a photoionized plasma with high column densities
suppressing resonant diffusion \citep{coupe04,godet04}. From the numbers
in Table~\ref{elines} we compute $G=2.2\,\pm\,1.2$ for O\,{\sc vii}
and $G=3.3\,\pm\,2.0$ for N\,{\sc vi}, both greater than one, but
with large uncertainties.

 At low densities the ratio of $R=$f/i is expected to be 3.9 for
O\,{\sc vii} and 4.9 for N\,{\sc vi} (low-density limit $R_0$),
while in a high-density plasma, this ratio
is reduced by collisional excitations from the upper level of
the f-line into the upper level of the i-line, followed by
radiative de-excitations from there into the ground. We measured
f/i ratios of $1.1\,\pm\,0.1$ and $2.8\,\pm\,0.2$, which correspond to
densities of $\log(n_e)=11$ and $\log(n_e)=10.8$ for O\,{\sc vii} and
N\,{\sc vi}, respectively. We used the summarized parameterized
form given by, e.g., equation 2 in \cite{denspaper} that has been
derived by \cite{gj69} with the critical densities
$\log(N_c)=10.5$ and $\log(N_c)=9.7$ for O\,{\sc vii} and
N\,{\sc vi}, respectively. As a first approach, we neglected the
radiation term $\phi/\phi_c$. For comparison, \cite{nye07} measured
densities of order $10^5-10^{10}$\,cm$^{-3}$ from
various infrared lines between days 64 and 84, and our values
are rather high compared to that. However, the IR and the X-ray
emission may originate from different places in the flow,
especially if the X-ray lines originate from the shock (see
previous section).

 We next study the possibility that UV radiation fields lead
to the low f/i ratios. \cite{ness_CS} have given a method that
converts measured UV fluxes at the wavelengths $\sim 1630$\,\AA\
for O\,{\sc vii} and at $\sim 1900$\,\AA\ for N\,{\sc vi} into
the parameter $\phi/\phi_c$ that we have neglected above
\citep[see also][]{blum72,waldcas01}. This method includes a
dilution factor which can be converted into the distance between
the plasma and the UV radiation source, assuming a distance to 
RS\,Oph of 1.6\,kpc. We examined the simultaneous UVOT Grism data
taken with \swift, but the data are not calibrated yet, and the
wavelength range around 1630\,\AA\ is not covered.
We, therefore, inspected the archive of the  International
Ultraviolet Explorer satellite (IUE), which contains a number of
observations of RS\,Oph taken after the outburst on 1985, January 28
\citep[e.g.][]{shore96}. We extracted an observation taken 1985,
Feb. 26, 23:47:00UT ($\sim 30$ days after outburst; ObsID SWP25327)
and estimated flux levels for the continuum of
$\sim 2\times 10^{-13}$\,erg\,cm$^{-2}$\,sec$^{-1}$ at 1630\,\AA\
and 1900\,\AA. Emission lines near these wavelengths from
He\,{\sc ii} (1640\,\AA), Si\,{\sc iii}] (1892\,\AA), and C\,{\sc iii}]
\citep[1908\,\AA;][]{shore96} can increase the relevant flux to no more than
$10^{-12}$\,erg\,cm$^{-2}$\,sec$^{-1}$. The former flux level
leads to a distance between the UV source and the immersed
plasma of $\sim 4\times 10^{11}$\,cm (6\,R$_\odot$) for O\,{\sc vii}
and $\sim 3\times 10^{12}$\,cm (40\,R$_\odot$) for N\,{\sc vi},
while the latter flux leads to $\sim 9\times 10^{11}$\,cm
(13\,R$_\odot$) and $\sim 10^{13}$\,cm (180\,R$_\odot$) for
O\,{\sc vii} and N\,{\sc vi}, respectively.
These numbers are quite uncertain because we assume that the UV
emission level was similar during the 1985 and 2006 outbursts.
However, we can conclude that if the plasma emitting
these lines is further away from the UV radiation source,
than 200\,R$_\odot$, then the density must be higher than
10$^{10}$\,cm$^{-3}$. This distance is small compared to the
extent of the expanding shell on day 39.7. With an expansion
velocity of 1100\,km\,s$^{-1}$ the shell has reached a radius of
$3.8\times10^{14}$\,cm (5400\,R$_\odot$).

\subsection{Spectral Variability from High- to Low-flux phases}
\label{variability}

\begin{figure}[!ht]
\resizebox{\hsize}{!}{\includegraphics{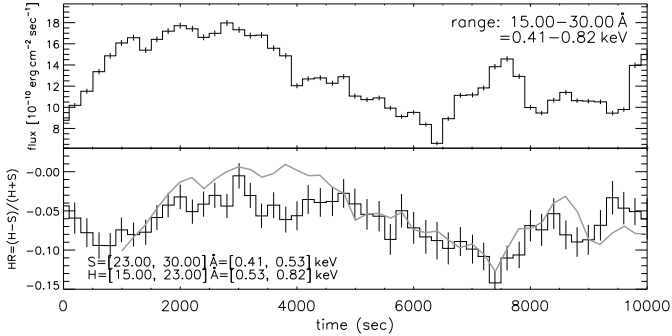}}
\caption{\label{hr}{\bf Top}: \chandra\ light curve extracted from the
dispersed photons in the wavelength range 15--30\,\AA\ (day 39.7).
{\bf Bottom}: evolution of hardness ratio (in steps of 200\,sec)
defined by the energy ranges given in the legend. The grey curve
is the light curve from the top panel, shifted by 1000\,sec on the
time axis and rescaled with a factor 0.012.}
\end{figure}

 We now focus on spectral changes on shorter time
scales. During the observation taken on day 39.7, variations
in brightness were detected which were not seen in the other two
observations. In Fig.~\ref{hr} we show the evolution of the observed
flux integrated over the wavelength range 15--30\,\AA\ (top panel) in
comparison to the hardness ratio HR (bottom panel) with the same
definitions as used in Sect.~\ref{spec} (repeated in the legend).
The rescaled light curve, marked
with the grey curve in the bottom panel of Fig.~\ref{hr} indicates that
the hardness ratio has the same shape but is retarded by 1000\,sec.

\begin{figure}[!ht]
\resizebox{\hsize}{!}{\includegraphics{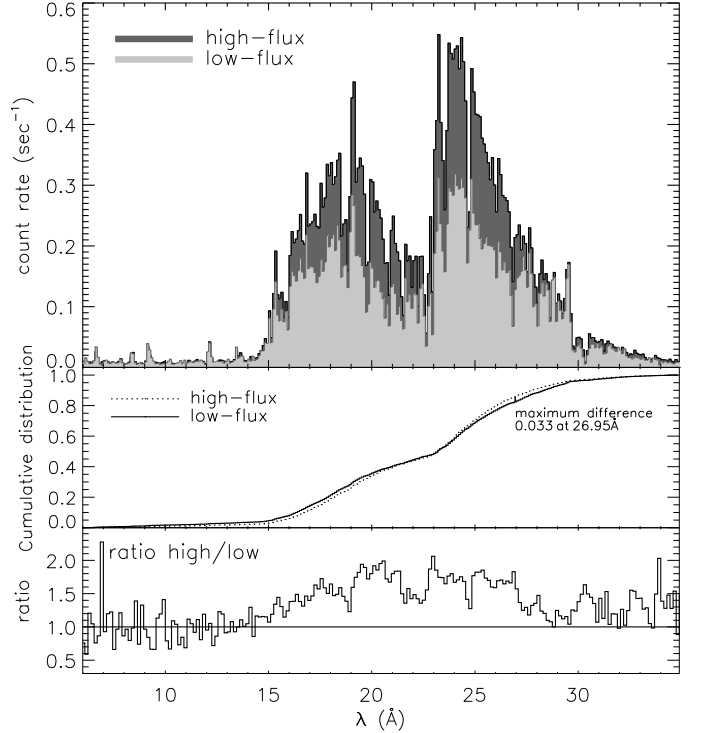}}
\caption{\label{tres}Comparison of spectra extracted from high-flux
and low-flux intervals in ObsID 7296. {\bf Top}: direct comparison of
count rate spectra. {\bf Middle}: Cumulative count distribution.
{\bf Bottom}: Ratio of spectra.}
\end{figure}

 In order to study the spectral changes in more detail, we extracted
spectra from the two different time intervals marked in dark and light
shadings in Fig.~\ref{lc}, representing high-flux
and low-flux emission. In Fig.~\ref{tres} we compare these spectra
with the dark shading being the high-flux spectrum
and the light shading the low-flux spectrum. In the bottom two panels we
show the cumulative distribution of counts and the ratio spectrum,
respectively.

 The cumulative distribution traces the spectral shape regardless of the
different intensities and can be used for a Kolmogorov-Smirnov 2-sample
test. At $\sim 27$\,\AA\ we find the largest difference of 0.03, which
has to be compared to the maximally allowed difference of 0.007 if the
two spectra were to be considered identical within 95\,percent
probability. The spectra are therefore different in their shapes in
addition the obvious difference of intensity. Inspection of the
cumulative curve reveals that the high-flux spectrum is slightly softer.

 The ratio spectrum shows up to a factor of two higher emission in the
high-flux spectrum. The absorption lines of O\,{\sc viii} at 19\,\AA\
and N\,{\sc vii} at 25\,\AA\ as well as the O\,{\sc i} absorption edge
at 22.83\,\AA\ are deeper relative to the continuum in the high-flux
spectrum. This indicates that the brightness changes involve more than
changes in the brightness of the continuum. In contrast,
shortwards of 15\,\AA\ as well as near 29.5\,\AA\ (the wavelength of
the N\,{\sc vi} forbidden line) the two spectra are identical.

\begin{figure}[!ht]
\resizebox{\hsize}{!}{\includegraphics{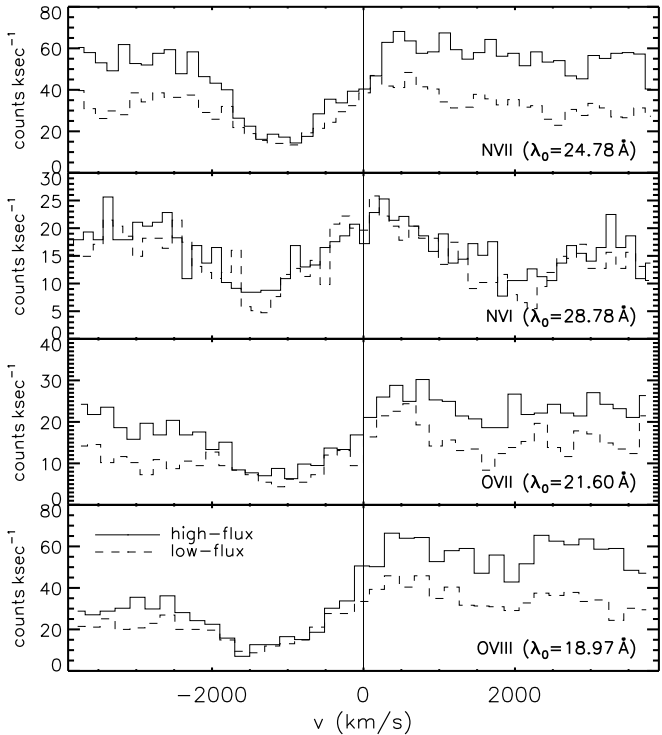}}
\caption{\label{lshape1_7296_lh}Details of \chandra\ spectra of day
39.7 for high-flux (solid) and low-flux (dashed) phases, plotted in
velocity space for the lines indicated in the legends of each panel.
The absorption lines are blue shifted by the same amount in both spectra.}
\end{figure}

 In Fig.~\ref{lshape1_7296_lh} we compare the high-flux and low-flux spectra
around the absorption lines in velocity space. While the high-flux phase
provides additional continuum emission, the emission level
in the troughs of the absorption lines is unchanged. There is no detectable
difference in blue shift of the absorption troughs.

We repeated our model from Sect.~\ref{pcygsect} for these two spectra. While
the optical depths at line center were around unity for both spectra, the
line column densities were significantly lower in the low-flux observation.
For example, for O\,{\sc viii} we found a column density of
$6.33^{+1.58}_{-1.51}$ in the high-flux spectrum but only
$2.93^{+1.92}_{-1.16}$ in the low-flux spectrum. The line shift velocities
were the same in the models to both spectra.

\section{Discussion}
\label{disc}

 In Sect.~\ref{anal} we have presented a number of issues that can
be addressed with the three grating spectra of RS\,Oph. The shape of the
spectrum cannot be approximated by a blackbody model. The dominant
feature, the O\,{\sc i} absorption edge, changes with time. The
quantitative analysis of these changes is complicated by various
overlapping absorption processes. For example, the
ionization edges and 1s-2p transitions from oxygen in low ionization
stages introduce significant structure to the observed spectrum.
These low ionization stages can occur in both the circumstellar and
the interstellar material in the line of sight. We thus need a more
refined model accounting for more processes than only K-shell
ionization of neutral elements in order to understand the effects
that the interstellar plus circumstellar material have on the observed
spectra.

 These uncertainties affect the interpretation of the origin and
production mechanism of the soft emission lines listed in the bottom
part of Table~\ref{elines}. While the hard lines at short wavelengths
can only be collisionally excited and originate from the shock, the
soft lines can in addition be photoexcited because of the presence
of the bright SSS spectrum. The contribution from photoexcitations
depends on the distance between the hot WD atmosphere that provides
the radiation and the plasma that emits the soft lines. If they
originate from the shock, this distance will be greater than if
they are part of the hot WD atmosphere. In the latter case
the soft lines should be much stronger than in the former case.
However, the amount of assumed extinction will be enhanced by
circumstellar material in addition to the interstellar medium,
leading to lower line fluxes measured at Earth.
In order to solve the ambiguity from these competing processes,
an improved absorption model for interstellar and circumstellar
absorption is required to derive accurate unabsorbed lines fluxes
for each case. The hard lines can be used to constrain the mean
emission measure distribution for a collisional plasma as developed by
\cite{ness_vel} for the Classical Nova V382\,Vel. This will be part
of an upcoming paper (Ness et al. in prep). If the soft lines originate
from the shock, they will show up with an excess in the emission
measure distribution, reflecting the contributions from
photoexcitations. The amount of excess depends on the distance between
the plasma and the hot WD atmosphere, which has to be of the same
order as that between atmosphere and the shocked plasma.
We note that the disappearance of the soft emission lines between
days 39.7 and 66.9 is consistent with the fading of the lines produced
in the shock. The only soft lines measurable on day 66.9 were also
weaker than on day 39.7, even though the SSS spectrum was brighter.
We thus favor a scenario in which the soft lines originate from the
shock and are partially photoexcited, while being absorbed only
by interstellar material.

 The optical depths at line center and the
velocities derived from the absorption lines (Sect.~\ref{pcygsect})
are not affected by the uncertainties in interstellar and/or
circumstellar absorption. The reduction of the velocities is due
to observing deeper into the flow at the time of
the later observation.

 Our absorption line model can be improved beyond the results
presented in Table~\ref{tab_profiles}. First, we need to
include the instrumental line profile in order to explore the
line widths. From our conclusion in Sect.~\ref{hesect} we can
define the boundary condition that the emission line components
are at their rest wavelengths, which leads to a reduction of
the uncertainties of the other parameters. Our above conclusion
that the soft emission lines likely originate from the shock has
to be accounted for by not allowing self absorption. We repeated
our models without self absorption and found similar results,
such that our results are robust.

 The issue of the appearance of P Cygni type profiles on day
39.7 cannot fully be answered at this stage. The shape of the
absorption lines is not typical for P Cygni profiles, which usually
show a sharp blue edge which represents the terminal velocity
of the wind. Our finding of symmetric absorption lines indicates
that we are not observing the entire velocity profile but
only a fraction of it. Those plasma regions that reside at the
terminal velocity (in velocity space) could be optically thin and
thus contribute little to the absorption lines. In the same
way, this plasma may also be too thin to produce measurable emission
lines, that would be scattered into our direction from the
photoexcited plasma moving perpendicularly to the line of sight,
further supporting our above conclusion that the soft lines
originate from the shock. Also, the He-like resonance lines are
expected to be significantly stronger than the intersystem lines
if photoexcitation of the circumstellar material was dominant.

 The high degree of variability during the first week of the SSS phase
(including our observation on day 39.7) is a new discovery that has
not been observed in any other nova. However, we have never had the
same frequency of observations of the same event. In one case, V4743\,Sgr
\citep{v4743}, a decline of X-ray brightness occurred, which
may be a similar phenomenon. Although this decay is only seen during
the first of four observations
taken during the SSS phase, it is not known when the SSS phase
started, and thus how early in the SSS evolution the decline was
observed. Variability was also observed in V1494\,Aql, V4743\,Sgr,
and V832\,Vel, but the amplitude was much lower and the variations
were periodic \citep[e.g.,][]{v4743,drake03}.

 In our \chandra\ observation on day 39.7 we discovered the hardness
ratio light curve lagging behind the total light curve by 1000\,s.
The hardness ratio is commonly used as a temperature indicator.
If the temperature varies, then this leads to an
increase in brightness before the spectral hardness increases.
Temperature variations could occur periodically, but we did not
detect any periods in the observation on day 39.7. With
the existing observations, we cannot exclude the presence of any
periods longer than three hours or shorter than a few days. However,
photoionization of oxygen could also lead to the higher hardness
ratios, as ionized oxygen allows more hard emission to pass through.
In this case the material in the line of sight responds
to the brightness of the continuum emission by adjusting the
degree of ionization within 1000 s. This timescale would
be the same for ionization and recombination and implies
a density of the absorbing material of $\sim 10^{10}-10{11}$\,cm$^{-3}$. 
This is consistent with the densities derived from the He-like
triplets, and the changes in hardness can therefore be explained
by variations in the degree of ionization of the circumstellar
material. However, this does not explain the origin of the
brightness changes.

 The spectra extracted from two different time intervals during the
episode of high-amplitude variations show complex changes. We found
deeper absorption lines and a deeper oxygen edge during times of
brighter continuum emission. This shows that the brightness
variations are correlated to the absorption behavior of the
expanding shell and the surrounding material and each spectrum
must be treated independently.

As an example, we applied our model from Sect.~\ref{pcygsect}
to the two spectra presented in Sect.~\ref{variability} and found
higher column densities during the high-flux phase while the
velocities were the same for both spectra.
Higher column densities arise when we either look deeper into the
outflow (longer path length) or when the density is higher (or
both). The high-flux phases can have their origin in a
reduction of the opacity, allowing us to view deeper into the
outflow where more emission is produced, the densities are higher,
and the temperatures are higher.

 Most of our results are based on preliminary models, and we are
improving these models. The first steps will
have to improve the atmosphere and the model for interstellar and
circumstellar absorption. The stellar atmosphere has to be modeled as
an expanding atmosphere with PHOENIX
\citep[e.g.][]{petz05}. The O edge has not been explored much, and
new models have to be developed \citep[see, e.g.,][]{paerels01_abs}.
We will further improve our model that fits the absorption and emission
lines simultaneously (Sch\"onrich et al. in prep). With improved models
in place, the changes of temperatures, luminosities and absorption
behavior during the phase of extreme variability can be determined.
A separate analysis of the emission lines and continuum produced in
the shocked plasma is under way (Ness et al. in prep), and more
detailed hydrodynamic models are being developed for comparison with
the data \citep[see e.g.][]{vaytet07}. Finally, the
\swift\ spectra can be used to constrain the evolution between the
grating observations. The model parameters obtained from grating
spectra on days 37.9, 54.0, and 66.9 can be interpolated for days
on which \swift\ spectra are available. The spectral models obtained
from the interpolated parameters have to agree with the respective
measured \swift\ spectra.

\section{Summary and Conclusions}
\label{concl}

The X-ray grating spectra of the SSS phase are complex and contain
a great deal of information. The objective of this paper is to provide 
an overview of the physical information that can be obtained from 
X-ray data. We have found:

\begin{itemize}
\item The brightness changes between the three observations are
consistent with the variations measured from the \swift\ 
observations at the same epochs.
\item Short-period oscillations ($\sim$ 35\,sec), first detected in some
of the \swift\ observations, are confirmed. We discovered that this
period resides in the plasma that emits the SSS spectrum.
\item The episode of high-amplitude variations detected with \swift\
that occurs on time scales of days is accompanied by variations on
time scales of hours.
\item The depth of the O\,{\sc i} absorption edge is variable, yielding
the lowest optical depth on day 54.0. One possible explanation is
photoionization of the absorbing material in the line of sight.
\item The absorption lines are shifted by \vaverageone\ on day 39.7 and
\vaveragetwo\ on day 66.9.
\item Declining emission from the shock is observed in the form
of collisionally excited emission lines shortwards of the SSS continuum.
These lines are slightly blue-shifted.
\item Superimposed on the SSS spectrum are emission lines
whose origin could either be from the shocked plasma or from photoexcited
material in the outer regions of the outflow.
\item Intersystem lines from He-like ions, are found
at their rest wavelengths. The ratio (i+f)/r (r the
1s-2p resonance line) is greater than one, indicating contributions from
recombination into excited states and small contributions from
resonance scattering. The ratio f/i indicates either densities in
excess of $10^{11}$\,cm$^{-3}$ or a short distance between the
plasma containing the He-like ions and a source for UV radiation.
\item The brightness changes during the phase of variability induce
changes in spectral hardness, retarded by 1000\,sec. Absorption
lines and the O edge are deeper during high-state phases, and
if the density of the absorbing material is of order
$10^{10}-10^{11}$\,cm$^{-3}$, then the hardness changes can
exclusively be explained by ionization. In that case the increases
and reduction of brightness occur underneath the absorbing
material, possibly in the regions where nuclear burning takes
place.

\end{itemize}

\acknowledgments

JUN gratefully acknowledges support provided by NASA through \chandra\ Postdoctoral
Fellowship grant PF5-60039 awarded by the \chandra\ X-ray Center, which is operated by
the Smithsonian Astrophysical Observatory for NASA under contract NAS8-03060.
SS received partial support from NSF and NASA grants to ASU.
PHH was supported in part by DFG grant HA 3457/2 and by the P\^ole
Scientifique de Mod\'elisation Num\'erique at ENS-Lyon.
KLP, JPO, APB, and MRG acknowledge support by the Particle Physics and
Astronomy Research Council. MFB is grateful to the UK
PPARC for the provision of a Senior Fellowship.
RS acknowledges support from Stiftung Maximilianeum,
Studienstiftung des deutschen Volkes, and the Max-Weber-Programm.
We are grateful to Harvey Tananbaum and the Chandra
Observatory for a generous allotment of Directors Discretionary.
Some of the observations have been obtained with XMM-Newton, an ESA science
mission with instruments and contributions directly funded by ESA Member States and NASA.
Some preliminary calculations carried out for this project were performed at the
H\"ochstleistungs Rechenzentrum Nord (HLRN), and at the National Energy
Research Supercomputer Center (NERSC), supported by the U.S. DOE, and  
at the computer clusters of the Hamburger Sternwarte, supported by the DFG  
and the State of Hamburg. We thank all these institutions for a generous
allocation of computer time. We thank an anonymous referee for
instructive comments.

\bibliographystyle{apj}
\bibliography{rsoph,cn,astron,jn}

\end{document}